\documentclass[final,3p,times,12pt]{elsarticle}

\usepackage{subcaption}
\usepackage{graphicx} 

\usepackage{lipsum}
\makeatletter
\def\ps@pprintTitle{%
	\let\@oddhead\@empty
	\let\@evenhead\@empty
	\def\@oddfoot{}%
	\let\@evenfoot\@oddfoot}
\makeatother

\usepackage{amssymb}
\usepackage{amsmath}
\usepackage{amsthm}
\usepackage[linesnumbered,ruled]{algorithm2e}

\usepackage{setspace}
\newcounter{bla}

\newcommand{\vek}[1]{\mathbf{#1}}
\newcommand{\nif}{\hat{\mathbf{n}}_s}

\begin{document}

\begin{frontmatter}

\title{TwoPhaseFlow: An OpenFOAM based framework for development of two phase flow solvers}

\author[a]{Henning Scheufler \corref{author}}
\author[b]{Johan Roenby}

\cortext[author] {Corresponding author.\\\textit{E-mail address:} Henning.Scheufler@dlr.de}
\address[a]{DLR Institute for Space Systems, Bremen}
\address[b]{Department of Mathematical Sciences, Aalborg University, Copenhagen}

\begin{abstract}
\begin{singlespacing}
One of the prevailing challenges in Computational Fluid Dynamics is accurate simulation of two-phase flows involving heat and mass transfer across the fluid interface. This is currently an active field of research, which is to some extend impaired by a lack of a common programming framework for implementing and testing new models. Here we present a new OpenFOAM based open-source framework allowing fast implementation and test of new phase change and surface tension force models. Capitalizing on the runtime-selection mechanism in OpenFOAM, the new models can easily be selected and compared to analytical solutions and existing models. As a start, the framework includes the following curvature calculation methods for surface tension: height function, parabolic fit, and reconstructed distance function method.  As for phase change, interface heat resistance and direct heat flux models are available. These can be combined with three solvers covering the range from isothermal, incompressible flow to non-isothermal, compressible flow with conjugated heat transfer. By design, addition of new models and solvers is straightforward and users are invited to contribute their specific models, solvers, and validation cases to the library.
\end{singlespacing}
\end{abstract}


\end{frontmatter}

\newpage
\begin{small}
{\bf PROGRAM SUMMARY}

\noindent
{\em Program Title: TwoPhaseFlow Library}                                   \\
{\em Licensing provisions: GNU General Public License 3 (GPLv3) }           \\
{\em Programming language: C++ (OpenFOAM v1812)}                            \\
{\em Nature of problem:}                                                    \\
Phase change is encountered in numerous applications and numerous numerical approaches exist to simulate this phenomenon. The main challenges in the direct simulation of these phenomena are parasitic currents caused by the surface tension model, the evaluation of the phase change mass, and the velocity jump condition at the interface. This library addresses these challenges and offers a framework with multiple surface tension and phase change models that simplifies the implementation and verification of new models. \\
{\em Program obtainable from:} \\
https://github.com/DLR-RY/TwoPhaseFlow  \\
\end{small}

\section{Introduction}

In almost all engineering design processes involving fluid flows, simulations with Computational Fluid Dynamics (CFD) software play an ever bigger role as a complement, and some times even as a replacement, of empirical laws and physical experiments. A computationally very challenging subset of these problems is those involving multiple fluid phases with heat and mass transfer between them. Here models and numerical methods are still relatively immature and so the use of CFD for design optimisation is still limited within this arena. In this paper, we present a numerical modeling framework based on OpenFOAM that simplifies implementation and verification and thus enables faster development of more accurate models.


The formulation of the heat and mass transfer or surface tension force models is highly influenced by the numerical interface representation. With interface tracking methods, the implementation of the surface tension and mass transfer across the interface is relatively straightforward because the interface is represented directly as the mesh faces separating the two fluid regions. The main drawback of this approach is the difficulty in dealing with large topological changes of the interface. In interface capturing methods like Volume of Fluid (VoF) and Level-Set (LS) these challenges are handled automatically. However, the implementation of phase change and surface tension models is more challenging. The formulation of the surface tension or phase change models mainly depends on the interface representation which can be categorized as diffusive or sharp. Examples of diffusive interface methods are the Phase Field and the Colour function Volume of Fluid (CVoF) method. Examples of sharp interface methods are LS and geometric VoF. Currently, the geometric VoF method, isoAdvector, \cite{Roenby.2016} \cite{Scheufler.2019}  and a colour function VoF are available in the modelling framework and will form the basis of the implemented models for phase change and for surface tension. However, the framework makes it easy to implement other interface capturing models and combine them with available surface tension models and phase change models.

As mentioned, the implementation of the phase change model greatly depends on the interface representation. Hardt and Wondra \cite{Hardt.2008} presented an interface heat resistance model to be used with CVoF. It is based on the Schrage equation and applies the temperature source terms in the whole interface region. The mass source terms are not active directly in interface cells but only in the nearest neighbours to the interface cells in order to avoid interface smearing and numerical pressure oscillation. Nabil and Rattner \cite{Nabil.2016} released an open-source framework for the simulation of incompressible phase change phenomena with the addition of new surface tension models. The framework offers the implementation of interface heat resistance models which are coupled explicitly with the governing equations. Among other phenomena, the authors successfully simulated film condensation. Another class of phase change models is the direct heat flux (DHF) models, which Kunkelmann and Stephan \cite{Kunkelmann.2011} found to be more accurate at the same spatial resolution when compared to the model of Hardt and Wondra \cite{Hardt.2008}. These types of models require the geometrical information of the interface which was achieved by reconstructing the 0.5-isosurface of the volume fraction field provided by the colour function VoF model. The model assumes that the reconstructed interface is always at saturation temperature and computes the mass flux from the temperature gradient at the interface. As in Hardt and Wondra \cite{Hardt.2008} the mass source is numerically smeared over the interface region to avoid pressure oscillation. Batzdorf \cite{Batzdorf.2015} found that the explicit treatment of the source terms in the energy equation leads to a time step restriction and instability of the model. Batzdorf formulated the gradient calculation implicitly and successfully solved the time step limitation. Sato and Niceno \cite{Sato.2013} construct the interface in the same way as Kunkelmann and Batzdorf but modified the calculation of the distance between the cell centres and the interface position to achieve an implicit coupling. 

The accuracy of the phase change models depends on the precision of the temperature field which is influenced by spurious currents caused by numerical errors in the surface tension model \cite{Sato.2013}. Therefore, accurate simulations of small scale phase change phenomena require a precise prediction of the surface tension forces. Parasitic currents are caused by discretization errors of the pressure jump conditions at the free surface. The main challenge is to convert a force acting on a surface to a volumetric force. The pressure jump at the interface is proportional to the surface tension multiplied by the curvature. For a constant curvature e.g. a sphere, a well-balanced solution was found by Francois et al. \cite{Francois.2006}. Thus, an exact curvature model applied to a sphere would result in near machine precision velocities. However, computing a curvature that converges with mesh refinement has proven very difficult. Brackbill et al. \cite{BrackbillJeremiahU.DouglasB.KotheandCharlesZemach.1992} proposed what is probably the most well-known surface tension model calculating the curvature from the gradient of the volume fraction, but this results in a non-converging curvature and large spurious currents. A more accurate prediction of the curvature can be achieved with a geometric VoF by exploiting the interface reconstruction data. With a parabolic fit \cite{Popinet.2009} or with reconstructed distance function (RDF) \cite{Cummins.2005} the accuracy of the curvature calculation can be improved by more than an order of magnitude, however, still without convergence with mesh refinement. Currently, the most accurate approach is the height function method \cite{Cummins.2005} \cite{Popinet.2009} which is usually paired with geometric VoF. It achieves second-order convergence on the static reconstruction of a sphere or disc, yet in case of the pure advection of the sphere or disc it does not converge with mesh refinement \cite{Popinet.2009} \cite{Abadie.2015}. The spurious currents are well-known for surface tension force but can also arise from gravity forces as demonstrated by  Wroniszewski et al. \cite{Wroniszewski.2014}, where the choice of discretisation of the gravity term significantly affects the results. Here it is often convenient to incorporate the hydrostatic potential into the pressure because this leaves us with a gravity source term that is only active on the fluid interface. The modified pressure (sometimes referred to as the dynamic pressure) then has a jump at the interface similar to the pressure jump caused by surface tension when the interface curves.

This paper presents a coding framework incorporating multiple surface tension models and phase change models. The framework allows easy addition of new models and automated benchmark tests with common analytical functions. As will be demonstrated below, even without further modifications the models implemented in our framework significantly improve OpenFOAM's capability to simulate flows with surface tension and phase change phenomena compared to what is currently available in OpenFOAM (v2012).

\section{Governing equations}

An unstructured finite volume method is used to discretize the governing equations which are solved in a segregated approach. In the proposed framework, three solvers are available, with the most general solver, \texttt{multiRegionPhaseChangeFlow}, being a compressible non-isothermal solver accounting for the effects of phase change and conjugated heat transfer. The governing equations for this solver are given below.

The VoF method is used as the interface capturing method where the transport equation for the volume fraction is given by \cite{jadidi2014} \cite{westermaier2020} \cite{Koch2016} \cite{Miller.2013} 

\begin{equation}
\frac{\partial \alpha}{\partial t} + \nabla \cdot (\mathbf{u} \alpha) - \dot \alpha_{pc}  = \alpha \nabla \cdot \mathbf{u} + \alpha (1 - \alpha) \left(\frac{\psi^v}{\rho^v} -\frac{\psi^l}{\rho^l}\right)\frac{D p}{D t}.
\label{eq:alphaComp}
\end{equation}
Here $\alpha$ is the volume fraction\footnote{Strictly speaking the volume fraction is a cell averaged quantity, not a mathematical field, and so its usage in a differential equation is not properly defined. In finite volume literature it is, however, common practice to ignore this, and write partial differential equation as short-hand notation for their cell volume integrated counterpart.}, $\mathbf u$  is the velocity field, $p$ is the pressure, $\psi$ is the compressibility and, $\rho$ is the mass density. The superscript, $l$, denotes the liquid phase and $v$ the vapor phase. The two terms on the right-hand side account for the effects of volumetric changes due to heating or compression and can therefore be neglected in case of incompressible flow. The explicit source term, $\dot\alpha_{pc}$,  accounts for phase change. The first and second terms on the left-hand side are either discretized by the standard colour function VoF model or the geometric VoF model \cite{Roenby.2016} \cite{Scheufler.2019}.

The Navier-Stokes equations are written in the form
\begin{equation}
\frac{\partial ( \rho \mathbf{u})}{\partial t} + \nabla \cdot ( \rho \mathbf{u} \mathbf{u}) - \nabla \cdot \left\{ \mu_\textrm{eff} (\nabla \mathbf{u} + (\nabla \mathbf{u})^{T}) \right\} = -\nabla p_\textrm{rgh}  + (\mathbf g \cdot \mathbf  x) (\rho^l-\rho^v)\nif \delta_s   + \mathbf{f}.
\label{eq:NSE}
\end{equation}
Here we use the auxiliary quantity $p_\textrm{rgh}$ defined as
\begin{equation}
	p_\textrm{rgh} = p - (\mathbf{g} \cdot \mathbf{x})\rho
	\label{eq:p_rgh}
\end{equation}
where $\mathbf g$ is the gravity vector and $\mathbf x$ is the position vector, $\mu_\textrm{eff}$ is the effective viscosity, $\nif$ is the unit interface normal (pointing into the heavy fluid), $\delta_s$ is the Dirac delta function. Numerically, working with $p_\textrm{rgh}$ rather than $p$ has the advantages that specification of boundary conditions becomes simpler and that the remaining gravity term with the Dirac delta function is only nonzero at the fluid interface.  The final vector, $\mathbf{f}$, in Eqn.~\ref{eq:NSE} accounts for additional source terms such as surface tension. It should be noted that the effective viscosity in the case of laminar flow is identical to the dynamic viscosity. The velocity, $\mathbf{u}$, is represented by a single field representing the liquid (vapor) velocity at points in space occupied by liquid (vapor).  In the finite volume framework, where we average fields over a computational cell, the velocity field in a cell containing both liquid and vapor is thus a mixture of the local liquid and vapor velocity. Similarly, the cell volume average of the density, $\rho$, becomes the volume fraction weighted average density in cells containing both liquid and vapor.

The energy equation is formulated in terms of the temperature, $T^i$, in a two-field approach, where $i$ is either $v$ for vapor or $l$ for liquid,

\begin{equation}
\frac{\partial \alpha^i \rho^i c_p^i T^i}{\partial t} + \nabla \cdot ( \alpha^i \rho^i c_p^i T^i \mathbf u)= \nabla \cdot (\lambda^i \nabla T^i) + \dot{q}_{\mathrm{ph}}^i + \dot{q}^i. 
\label{eq:TLandTV}
\end{equation}
Here, $c_p$ is the specific heat and $\lambda$ is the thermal conductivity. The source term $\dot{q}_{\mathrm{ph}}$ accounts for energy  changes caused by phase change and is either an explicit or an implicit source term depending on the selected phase change model. Other effects such as e.g. compressibility effects of the gas phase are accounted for by the last term, $ \dot{q}$ . 

In the solid region, the heat transfer is calculated by
\begin{equation}
\frac{\partial \rho^s h^s}{\partial t} - \nabla \cdot \left(\frac{\lambda^s}{c_p^s}  \nabla h^s \right) = 0.
\label{eq:Tsolid}
\end{equation}
where $h^s$ is the enthalpy, $\rho^s$  the density, $c_p^s$ the heat capacity and $\lambda^s$ the thermal conductivity of the solid.

\section{Methods}
\sloppy
OpenFOAM  (Open-source Field Operation and Manipulation) is a finite volume framework written in C++ and with a focus on Computational Fluid Dynamics. Due to the generality and extensibility of the OpenFOAM framework a variety of applications can be addressed including chemical reactions, electromagnetics, structural mechanics, and heat transfer. One of the reasons for the huge success of the project is that the partial differential equations are easily recognizable as for example shown by the Navier--Stokes equations below:
\begin{footnotesize}
\begin{minipage}{\linewidth}
\begin{verbatim}
fvVectorMatrix UEqn
(
    fvm::ddt(rho, U) // rho -> density, U -> velocity
    // rhoPhi = rho*U.Sf: stored at the cell faces with area vectors Sf
    + fvm::div(rhoPhi, U) 
    + turbulence.divDevRhoReff(U) // accounts for diffusion of momentuum and selection of
    // mulitple turbulence models from the family of LES and RAS based models
    ==
    fvc::reconstruct // constructs the cell centre vector field
    (
        (	// terms below are defined on the faces for 
            // consistency with the pressure Poisson equation
            surfForces.surfaceTensionForce() // new surfaces force model
            + surfForces.accelerationForce(rho) // new acceleration force model
            // derivative normal to mesh faces of pressure minus hydrostatic potential
            - fvc::snGrad(p_rgh) 
        ) * mesh.magSf() // magnitude of cell face area vectors
    )
);
\end{verbatim}
\end{minipage}
\end{footnotesize}
where the first three lines correspond to the first three terms in Eqn.~\ref{eq:NSE}. The PDE's are stored in sparse matrices that employ two main discretisation techniques: implicit and explicit indicated by namespace fvm (finite volume method) and fvc (finite volume calculus). 

This software design enables the extension of existing solvers with additional source terms as shown above by the surface force module. All presented modules follow this design principle in which the governing equations for momentum, energy, and mass are extended by additional terms. The modules utilize an OpenFOAM specific runtime selection mechanism, enabling the selection of a model by changing a dictionary entry in the simulation case setup files. At the core of this mechanism is a virtual base class holding all relevant data and public functions. The derived classes implement the actual models which are then constructed in the top-level solver at runtime.

\fussy

\begin{figure}
	\centering
	\includegraphics[width=0.8\linewidth]{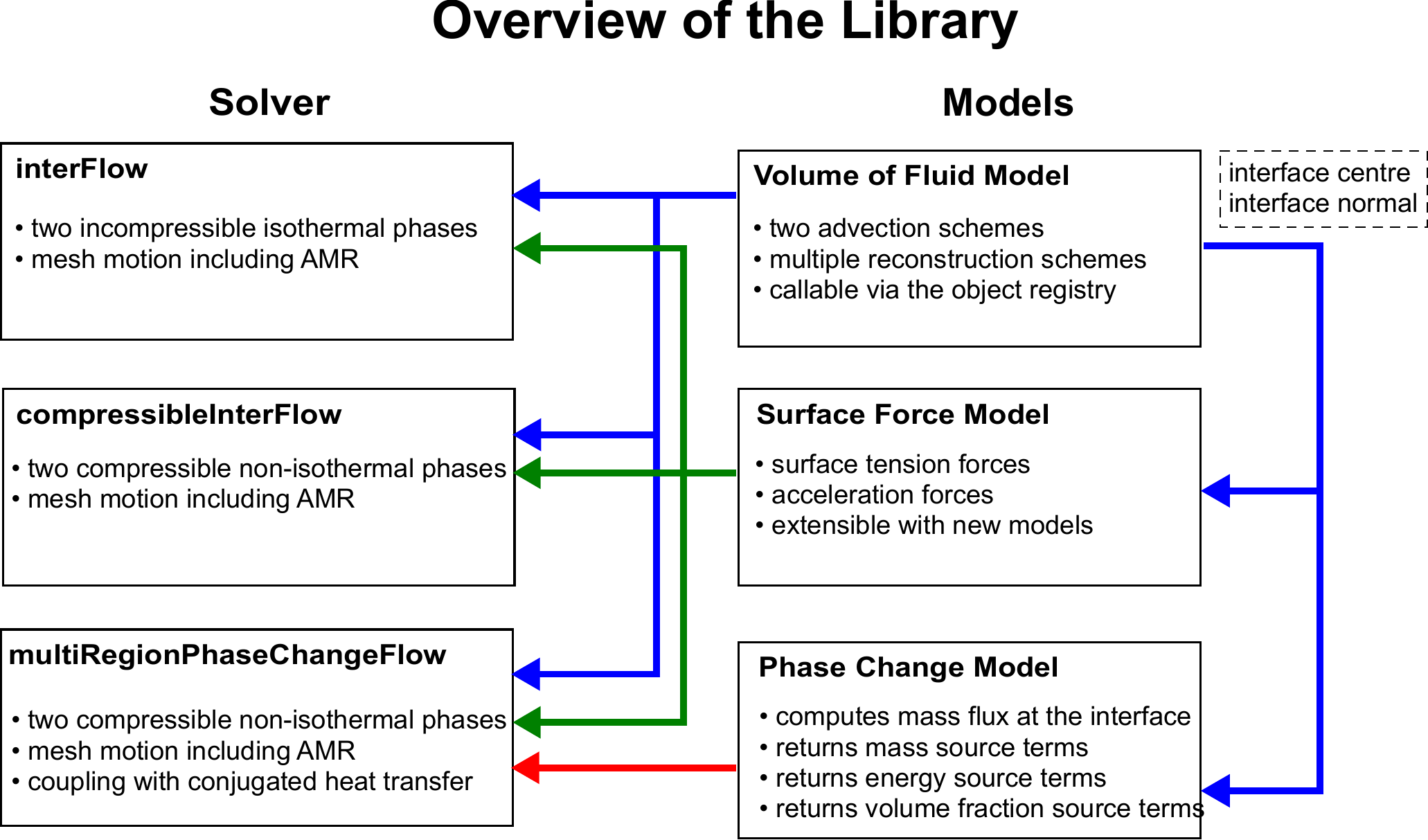}
	\caption{Schematic overview of the library}
	\label{fig:schematic_multiphaselib}
\end{figure}

Fig. \ref{fig:schematic_multiphaselib} shows an overview of the library, displaying three main modules and three solvers. The core of the library is the Volume of Fluid module which advects the volume fraction field, $\alpha$, and also reconstructs the interface inside each interface cell by calculating the polygon separating liquid and vapor inside the cell. The surface force and phase change modules utilize this data to compute e.g. curvature or phase change mass. This data is then provided to the top-level solvers as described above. An incompressible formulation without heat and mass transfer is available in interFlow. The compressibleInterFlow solver extends interFlow with heat transfer and a compressible formulation. Mass transfer and conjugated heat transfer are added to compressibleInterFlow in the solver multiRegionPhaseChangeFlow. All solvers are capable of utilizing automatic mesh refinement and mesh motion and use the volume of fluid method to represent the interface.

Implementation details about the models are found in the sections below.

\subsection{Volume of fluid module}\label{sec:VoF_module}

\begin{figure}
	\centering
	\includegraphics[width=0.8\linewidth]{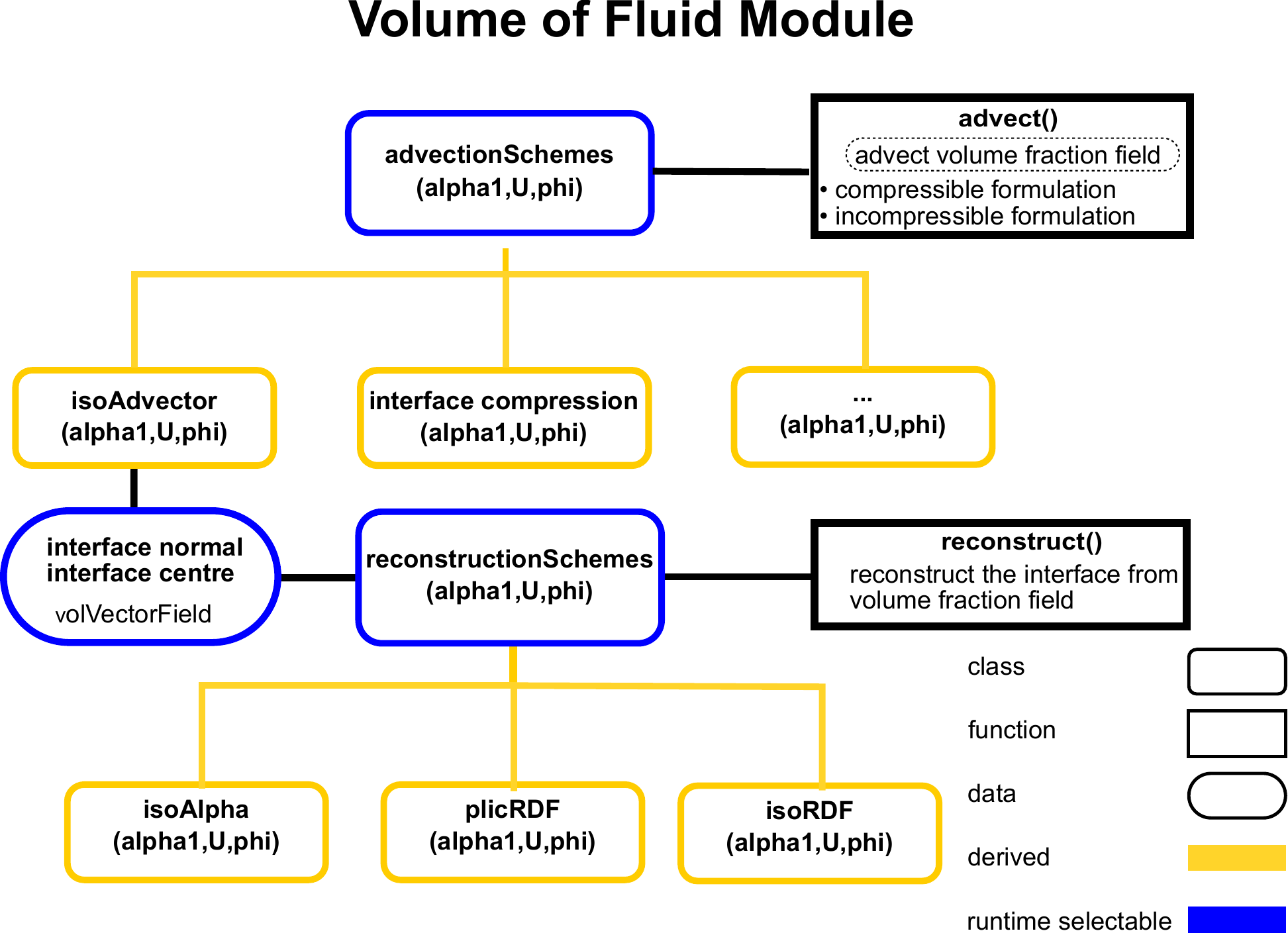}
	\caption{Schematic overview of the VoF module}
	\label{fig:schematic_VoFLib}
\end{figure}

The Volume of Fluid module is the core part of the library as it provides the interface reconstruction data for the other models and has already been released as open source by Scheufler and Roenby \cite{Scheufler.2019}. The library consists of two base classes: One for the reconstruction scheme and one for the advection scheme as illustrated in Fig. \ref{fig:schematic_VoFLib}. The advection scheme base class grants the possibility to integrate other advection schemes such as LS or phase field methods. However, currently only a colour function VoF model and the newly proposed geometric VoF method by Roenby et al. \cite{Roenby.2016} is implemented. The latter requires an interface reconstruction method to reconstruct the fluid interface from the volume fraction data. This is provided by the reconstruction scheme base class and its derived classes \texttt{isoAlpha}, and \texttt{plicRDF}. The \texttt{isoAlpha} model is identical with the model proposed by Roenby et al. \cite{Roenby.2016} and calculates the interface position and normal based on the proposed isosurface reconstruction method \cite{Roenby.2016}. \texttt{plicRDF} implements a PLIC (Piecewise Linear Interface Construction) scheme where the interface orientation is computed by the gradient of the reconstructed distance function (RDF). It achieves second order convergence on hexahedral, tetrahedral and polyhedral meshes as demonstrated in \cite{Scheufler.2019}.

\begin{figure}
	\centering
	\includegraphics[width=0.5\textwidth]{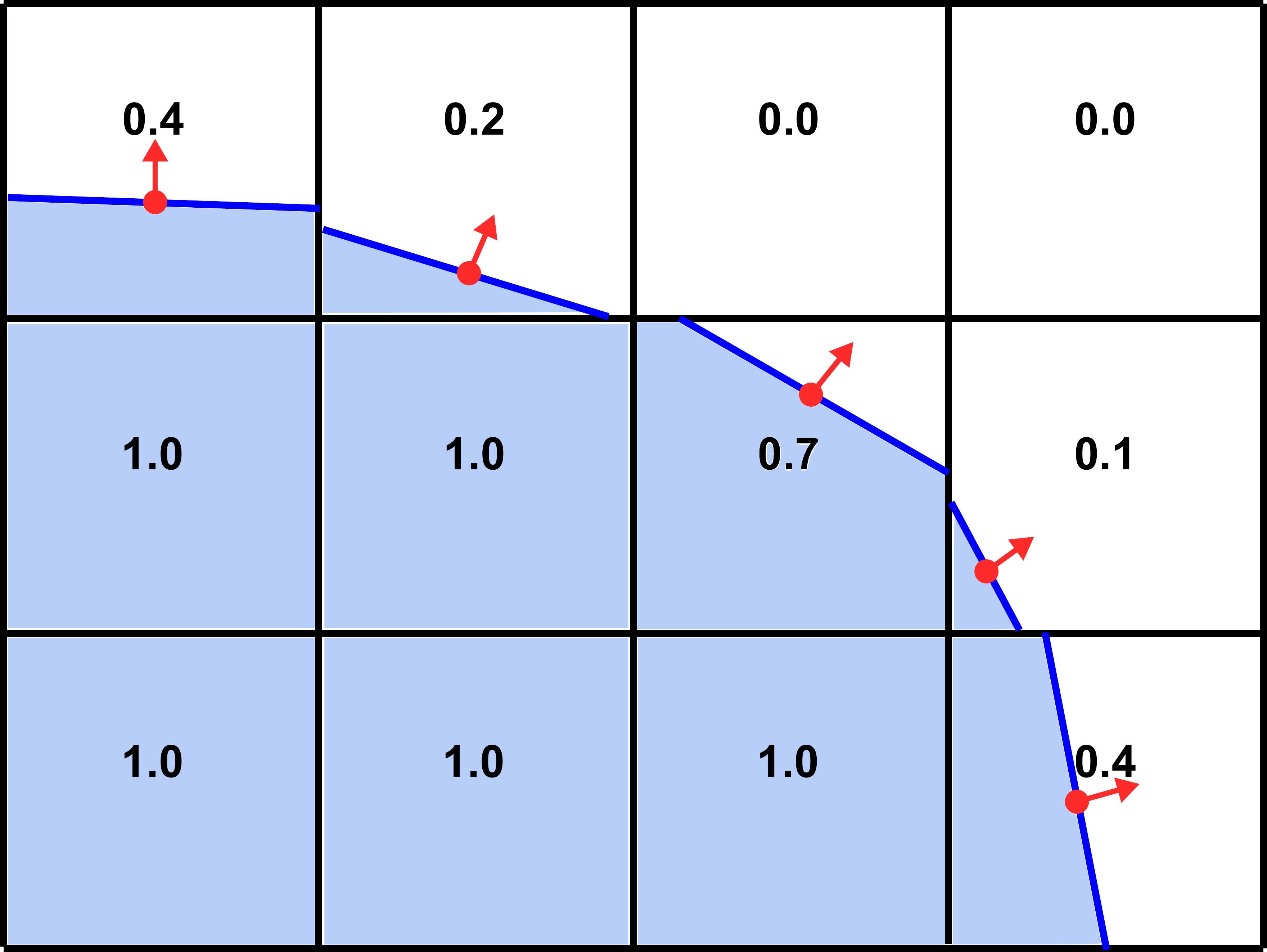}
	\caption{Example of a PLIC interface.}
	\label{fig:interface_reconstruction}
\end{figure}

Both reconstruction methods ensure that the interface segment inside a cell cuts the cell into subcells with volumes in accordance with the cell's volume fraction value. With linear/planar interface segements, this means that the interface lacks $C^0$ continuity as illustrated in Fig. \ref{fig:interface_reconstruction}. The PLIC and isosurface based reconstruction both compute the polygonal representation of the interface segment inside a cell. For each interface cell they store the interface centre point as well as the area vector of the interface segment. Each of these are stored in a so--called \texttt{volVectorField} which is an OpenFOAM class holding the value of vector fields in all cell centres (In cells not intersected by the fluid interface, the interface centre and area vector are set to the zero vector). These fields are required for the geometric VoF scheme but are also used in other submodules. The object registry that acts as OpenFOAM global database manages the pointer to the reconstruction scheme, granting access to the interface normal and position fields from every class. This enables the possibility to easily access the interface reconstruction data from anywhere in the solver. 

\begin{figure}
	\centering
	\includegraphics[width=0.5\textwidth]{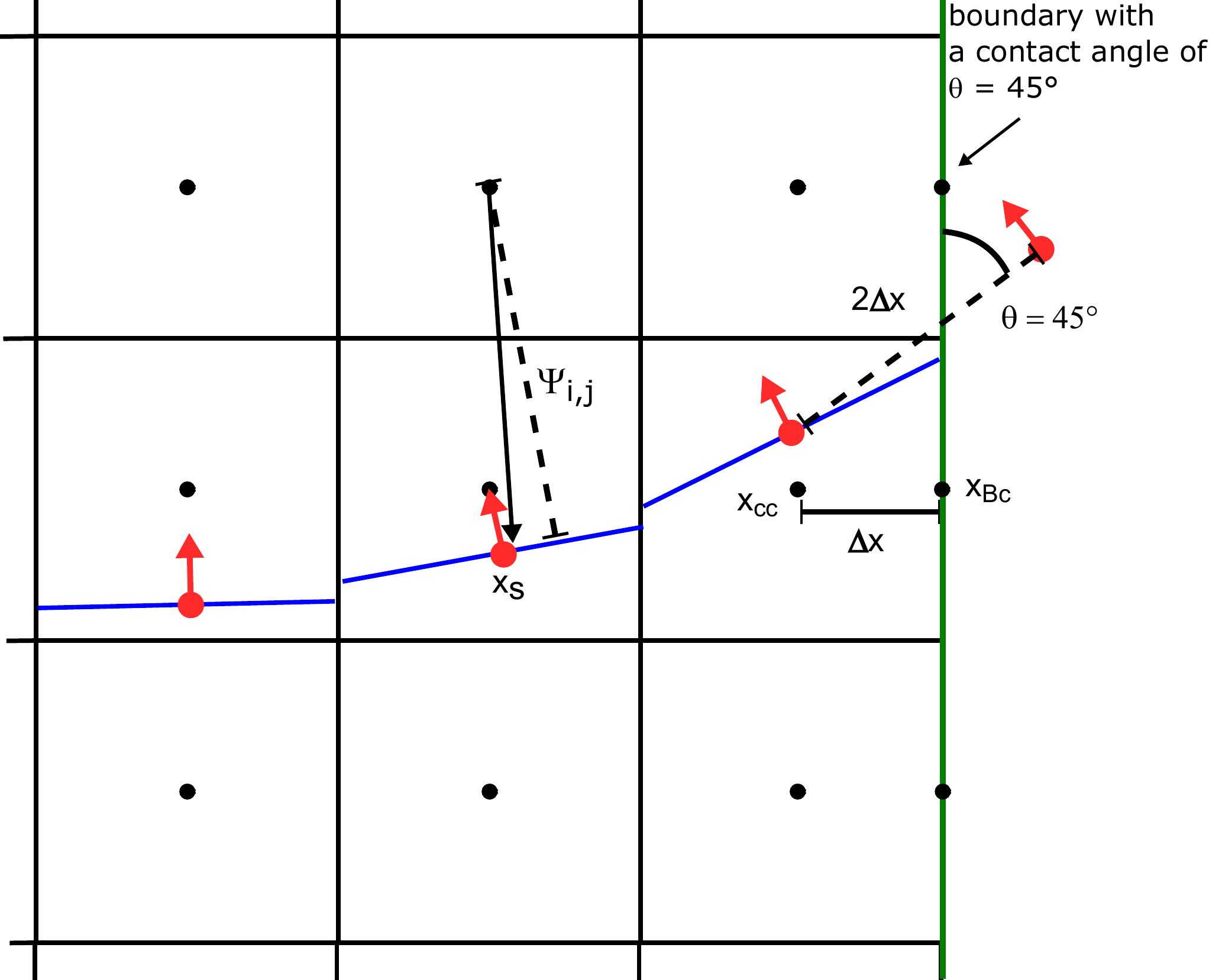}
	\caption{Illustration of idea behind extrapolation of the fluid interface to the boundary.}
	\label{fig:interface_BC_RDF}
\end{figure}

The surface tension module uses this information to compute (among other things) the curvature of the surface which relies on the accuracy of the interface reconstruction scheme. As shown in Scheufler and Roenby \cite{Scheufler.2019}, the interface position and normal can be accurately predicted but the prescribed contact angle on the boundaries, $\theta$, are not taken into account.  The proposed approach is still based on the RDF function that is defined as followed:
\begin{equation}
	\tilde \Psi_{cc,\vek x_s} =  \nif \cdot (\vek x_{cc} -  \vek x_s),
\end{equation}
where $\tilde \Psi_{cc,\vek x_s}$ denote distance from a cell centre to an interface segment. From these distances, the RDF in the centre of cell $cc$ is calculated as 
\begin{equation}\label{eq:cellCentrePsi}
	\Psi_{cc} =  \frac{\sum_{nei} w_{nei} \tilde \Psi_{cc,\vek x_s}}{\sum_{nei} w_{nei}},
\end{equation}
where the sum is over all point neighbours of cell $cc$ that are interface cells, and the weighting factor is chosen to be
\begin{equation}
	w_{nei} =  \frac{|\nif \cdot (\vek x_{cc} -  \vek x_s)|^2}{|\vek x_{cc} -  \vek x_s|^2}.
\end{equation}
The interface normal is then approximated with a least square fit as:
\begin{equation}\label{eq:LSN}
	\nif =  \nabla \Psi.
\end{equation}

Since the accuracy of the normal calculation only depends on the estimation of the RDF function, special treatment must be applied at boundary cells, where the user may want to specify a contact angle. For a boundary cell with a calculated interface centre, we propose to define a ghost interface point on the other side of the cell's boundary face. The position of the ghost interface point is then uniquely specified by choosing it to lie on the line passing through the boundary cell's interface centre and passing through the boundary face at the user specified contact angle and specified length. The orientation of the ghost interface normal is chosen so that its angle with the boundary normal equals the user specified contact angle. These choices of ghost interface point and orientation are illustrated in Fig. \ref{fig:interface_BC_RDF}. These ``ghost cell'' interface points and normals satisfying the contact angle requirement on the boundary are used in the weightings defining the value of the RDF function in the cell centres as determined by Eqn.~\ref{eq:cellCentrePsi}. This method for coping with contact angles is far from optimal but is our best current approach based on extensive numerical experimentation with various approaches. Generally, literature is very sparse with regards to interface reconstruction with the inclusion of boundary handling.

\subsection{Phase change module}
\begin{figure}
	\centering
	\includegraphics[width=0.8\linewidth]{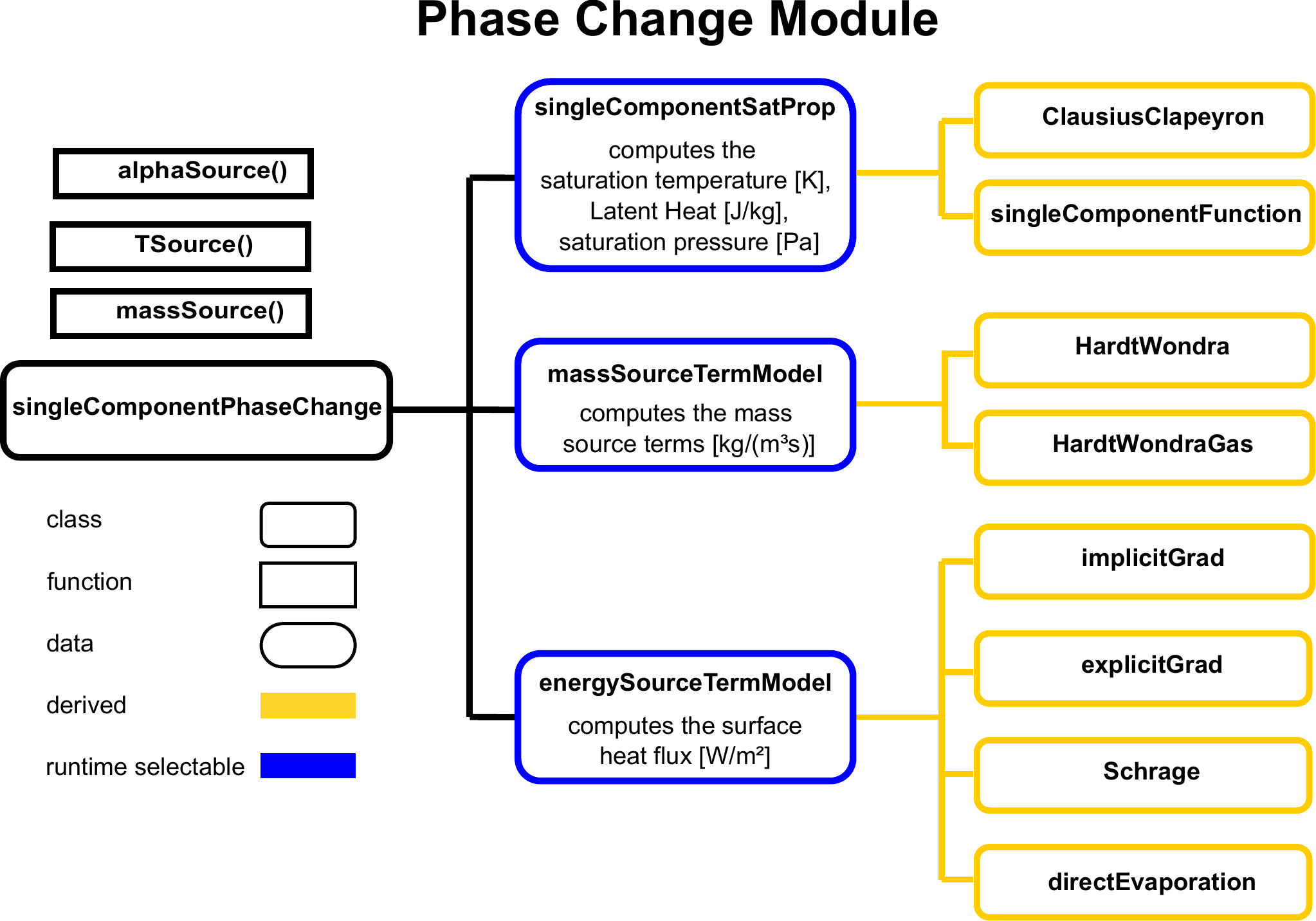}
	\caption{Schematic overview of the phase change module}
	\label{fig:schematic_phaseChangeModel}
\end{figure}

\sloppy
The phase change module computes the mass transfer at the liquid/gas interface in a one species system. The schematic overview of the module is shown in Fig. \ref{fig:schematic_phaseChangeModel} and is utilized in the custom solver \texttt{multiRegionPhaseChangeFlow} . The module consists of the wrapper class \texttt{singleComponentPhaseChange} that provides source terms for the volume of fluid, energy and mass equation. It wraps three runtime selectable classes implementing the specific models that are described in more detail below. \- 

\fussy

The \texttt{massSourceTermModel} computes the source terms for the VoF and mass equation while the \texttt{energySourceTermModel} computes the energy source terms. The material properties relevant for saturation are managed by a class called \texttt{singleComponentSatProp}. 

\subsubsection{Saturation properties}

The \texttt{singleComponentSatProp} class provides the saturation temperature, saturation pressure and heat of evaporation. Currently, two options to specify the above properties are available:

\begin{enumerate}
    \item Clausius--Clapeyron relation :
    \begin{equation}
        \ln(p/p_1) = -\frac{L}{R}\left(\frac{1}{T_1}-\frac{1}{T}\right)
    \end{equation}
    
    \item Functions of temperature and pressure:
    \begin{align}
     T_{Sat} & = f(p), \nonumber \\
     p_{Sat} & = g(T), \label{eqn:eqlabel} \\
     L & = h(p)\nonumber 
    \end{align}
\end{enumerate}

In the Clausius--Clapeyron relation, one must specify the latent heat, $L$, the specific gas constant, $R$, as well as a point, $(T_1, p_1)$ on the saturation curve in order to describe the saturation temperature and pressure. For a large pressure range, the assumption of a constant latent heat can be erroneous. Specifying the saturation properties as a polynomial function of pressure or temperature or interpolating it from a table is a more accurate approach, which is therefore also an option in the implemented code.

\subsubsection{Energy source terms}

The class, \texttt{energySourceTermModel}, is the core of the phase change library. It computes the energy source terms and provides it to the energy equation of the solver, Eqn.~\ref{eq:TLandTV}, by returning a matrix. Additionally, these source terms are also used in the mass source term model that (currently) smears and scales the provided source terms. In the literature, two types of energy source models are described \cite{Kunkelmann.2011}: 1) Interface heat resistance and 2) gradient-based models, both of which are implemented in the framework. All models compute a volume-specific power associated with phase change:
\begin{equation}
q_{pc} = \lambda^l \nabla T^l \cdot \nif + \lambda^v \nabla T^v \cdot (-\nif)
\label{eq:GradPHModel}
\end{equation}
for the gradient based models and by
\begin{equation}
q_{pc} = \frac{T^v - T_{Sat}}{R_{int}} + \frac{T^l - T_{Sat}}{R_{int}}
\end{equation}
\sloppy
for the interface heat resistance models, where $T_{Sat}$ is the saturation temperature. The models are implemented as derived classes of \texttt{energySourceTermModel} and compute heat transfer due to phase change. The values are then used to couple the solvers and to compute the phase change mass described in more detail below.
\fussy

\paragraph{explicitGrad}

\begin{figure}
	\centering
	\begin{minipage}{0.48\textwidth}
	\centering
	\includegraphics[width=0.9\linewidth]{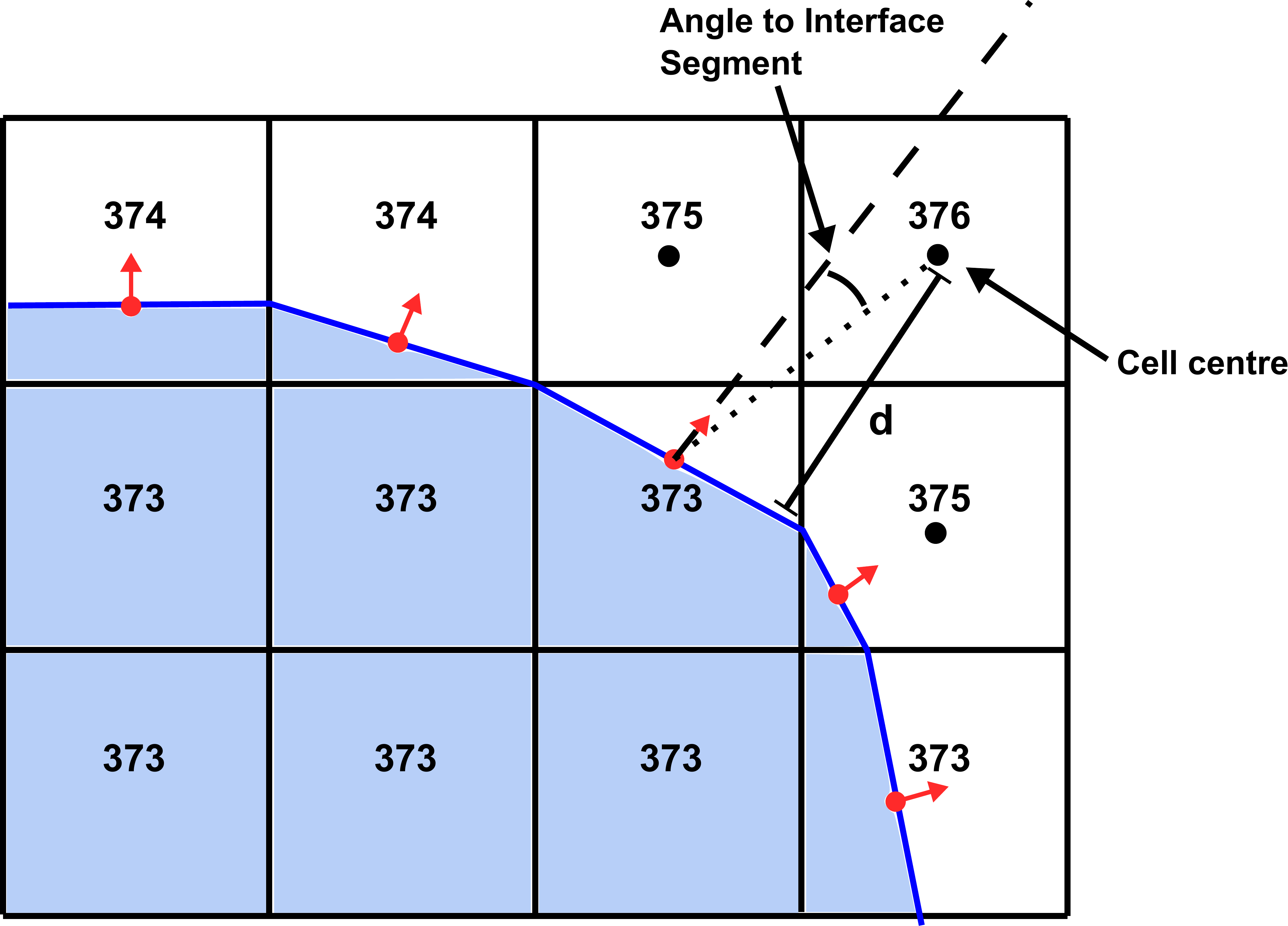}
	\caption{Gradient estimation of the explicitGrad model}
	\label{fig:explicitGrad}
	\end{minipage}\hfill
	\begin{minipage}{0.48\textwidth}
	\centering
	\includegraphics[width=0.9\linewidth]{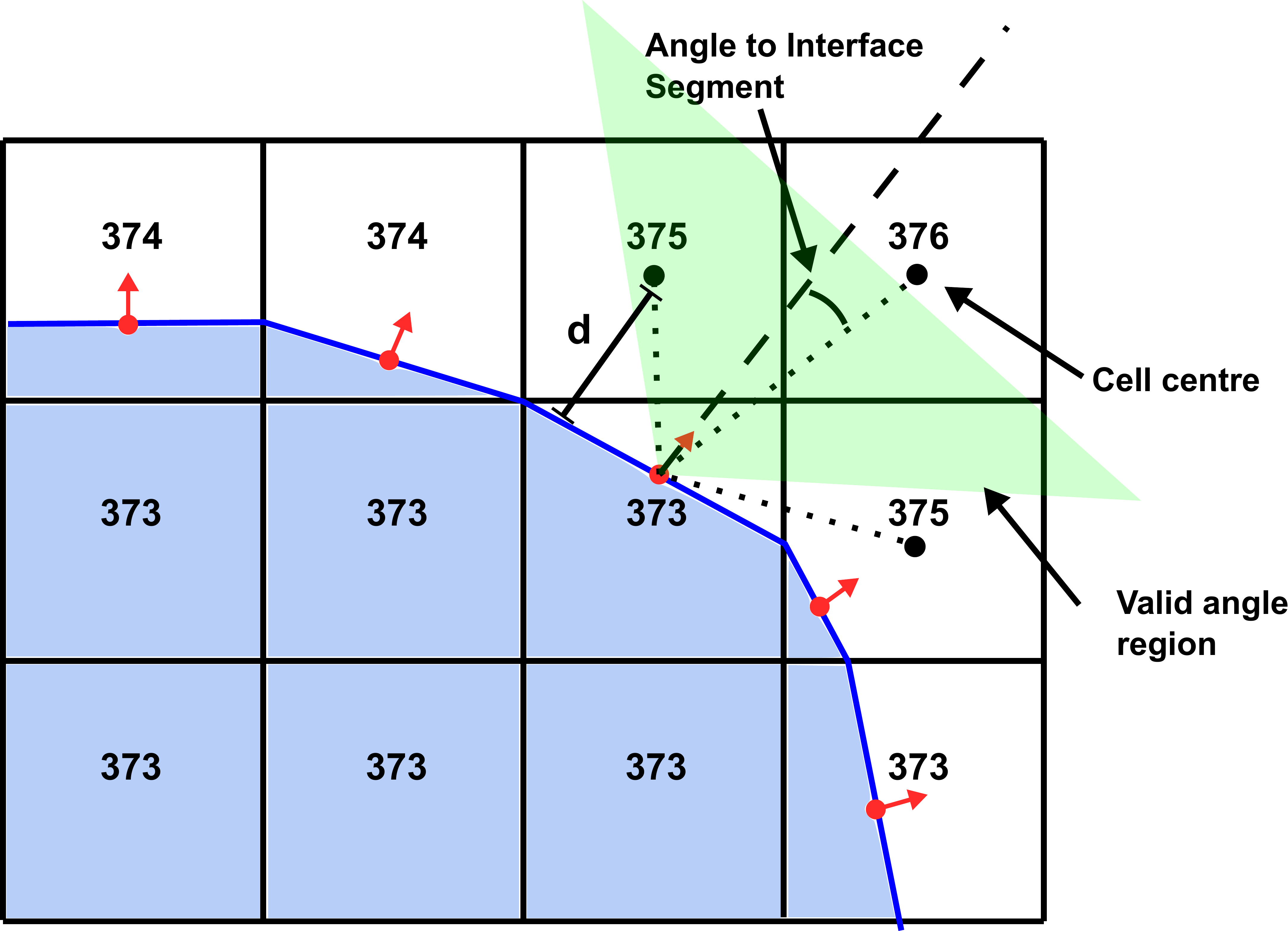}
	\caption{Gradient estimation of the implicitGrad model}
	\label{fig:implicitGrad}
	\end{minipage}
\end{figure}

This model is a simple implementation of a gradient-based model and is similar to the model described in Kunkelmann \cite{Kunkelmann.2011}. With the assumptions that the interface is on saturation temperature and that energy can only be transported by diffusion over the interface, the heat flux $q_{pc}$ can be computed by the Fourier law. The challenging part is the discretization of the temperature gradient at the interface depicted in Fig. \ref{fig:explicitGrad}.  For an interface cell with interface centre $\vek x_s$ and interface normal $\nif$ we chose to calculate the one-sided temperature gradient on the liquid side based on the temperature $T_\textrm{nei}$ in the neighbour cell towards which $\nif$ is pointing. That is, if a neighbour cell has centre $\vek x_\textrm{nei}$, then we choose the neighbour where $\vek x_\textrm{nei} - \vek x_s$ makes the smallest angle with $\nif$. From this choice we then approximate the temperature normal derivative as
\begin{equation}
\nif  \cdot \nabla T^l \approx \frac{T_\textrm{nei} - T_{Sat}}{\nif \cdot (\vek x_s - \vek x_\textrm{nei})}.
\label{eq:explicitGrad}
\end{equation}
The one-sided normal temperature gradient in the vapor phase, $\nif  \cdot \nabla T^v$ is calculated in the same way, but with the chosen neighbour cell in the vapor phase, i.e. the cell pointed at by $-\nif $ due to the convention that $\nif$ points out of the vapor region. 

With the normal gradients computed both on the vapor and liquid side of the fluid interface, we can calculate the heat flux using Eqn.~\ref{eq:GradPHModel}. The computed heat flux is multiplied by the interface area within a cell (provided by the volume of fluid module) and applied as explicit source term in the Eqn.~\ref{eq:TLandTV}. Due to the explicit nature of the source term, a stability criterion enforces a time step limitation on the solver.

\paragraph{implicitGrad}

To circumvent the time step constraint and stability problems that may arise with the explicit variant of the gradient-based models, Batzdorf \cite{Batzdorf.2015} proposed an implicit formulation which forms the basis for the \texttt{implicitGrad} model. The basic idea of this approach is to include part of the gradient on the diagonal of the matrix in the discretised energy equation,
\begin{equation}
q^i_\textrm{pc} = \sum_\textrm{nei} \frac{w_\textrm{nei}\lambda^i}{d_\textrm{nei}} T_{Sat} - \sum_\textrm{nei} \frac{w_\textrm{nei}\lambda^i}{d_\textrm{nei}}  T_\textrm{nei} \textrm{    with    } w_\textrm{nei} = \left( \frac{\cos \theta_\textrm{nei}}{\sum_{m}\cos \theta_m}  \right)^4.
\label{eq:implicitGrad}
\end{equation}
The angle $\theta_\textrm{nei} = \nif \cdot (\vek x_\textrm{nei}-\vek x_s)$, and for a given interface cell the sums are over all neighbour cells, where this angle is less than 70 degrees. Fig. \ref{fig:implicitGrad} illustrates this choice of neighbour cells. In contrast to the explicit variant, the source term is not applied directly in the interface cells but in the neighbouring cells. The second sum in Eqn.~\ref{eq:implicitGrad} is treated implicitly and hence adds to the diagonal of the matrix while the first sum is added to the source term of the matrix equation. The row of the matrix matches the cell index of the neighbouring cell. With this approach, no explicit time step criterion exists which greatly improves the stability of the solver.
 
\paragraph{Schrage}

The Schrage model is implemented as an interface heat resistance model. In this type of model the heat flux is computed from the temperature difference between the cell temperature, $T$, and the saturation temperature, $T_{Sat}$, as follows,
\begin{equation}
	q_{pc} = \frac{T^v - T_{Sat}}{R_{int}} + \frac{T^l - T_{Sat}}{R_{int}},
	\label{eq:Schrage}
\end{equation}
where the coefficient, $R_{int}$, is defined by
\begin{equation}
	R_{int} = \frac{2 -C_{acc}}{2 C_{acc}}\frac{T_{Sat}^{3/2}\sqrt{2 \pi R_{gas}}}{\rho^{v} L^2}.
\end{equation}
Here $C_{acc}$ is the accommodation factor, $R_{gas}$ is the specific gas constant and $L$ is the latent heat. As in the implicit gradient model, the first term of Eqn.~\ref{eq:Schrage} is added to the source of the matrix, while the second term adds to the diagonal of the matrix. This implicit treatment increases the stability of the solver. The difference between our implementation and the model proposed by Hardt and Wondra \cite{Hardt.2008} is in the calculation of the interface area where they use $|\nabla \alpha|$ while we use the geometrically calculated interface area provided by the VoF module. Furthermore, our implementation employs a temperature field for each equation, rather than a single field.

\paragraph{Direct evaporation}

The direct evaporation model is a combination of the interface heat resistance model and the gradient based model. This ``engineering model'' requires the specification of the superheated temperature and an interface heat resistance coefficient. If the temperature exceeds the superheated temperature the liquid is evaporated but with the assumption that the resulting volume increase instantly moves to the interface.

\subsection{Mass source terms}
Phase change causes a velocity jump at the interface which is modeled by applying source terms in the pressure Poisson equation. The magnitude of the velocity jump is proportional to the mass flux, $\dot{m}$, at the interface and can be calculated as
\begin{equation}
	\dot{m} = \frac{q_{pc}}{L}
\end{equation}
where the heat flux, $q_{pc}$, from the \texttt{energySourceTermModel} is divided by the latent heat. The resulting source term field results in a sharp source term distribution at the interface especially for the kind of geometrically reconstructed interface used here. Applying the resulting source term field directly at the interface results in pressure and velocity oscillation as well as a smearing of the interface \cite{Hardt.2008}. The implemented models circumvent the problem by smearing the source terms and by only applying them in the neighbourhood of the interface and not directly at the interface. The \texttt{massSourceTermModel} provides the source terms for the pressure equation and for the VoF equation.

\subsubsection{HardtWondra}

The Hardt and Wondra \cite{Hardt.2008} model avoids pressure oscillation and a smearing of the interface by smoothing the sharp source term distribution provided by the \texttt{energySourceTermModel}. A detailed description of the implementation can be found in Kunkelmann \cite{Kunkelmann.2011} and Batzdorf \cite{Batzdorf.2015}. In the first step, the sharp source term distribution is smeared with a Laplacian function,
\begin{equation}
    \rho_{smeared} - \Delta(D \rho_{smeared}) = \rho_{sharp},
\end{equation}
with the numerical diffusion coefficient defined by $D = (C \Delta x)^2$, $\Delta x$ being the cell size. The coefficient, $C$, is roughly the number of cells over which the interface is smeared (the default value is set to $C=3$). The second step is to set all source terms to zero in the interface region defined by: $\mathrm{cutOff} < \alpha < 1-\mathrm{cutOff}$ with the default value $\mathrm{cutOff}=1 \cdot 10^{-3}$. Laplacian smoothing keeps the volume integral of the source terms constant and is therefore a conservative operation. Obviously, setting the source terms to zero inside the interface region violates this conservation. Therefore, in the next step the source terms of the liquid and gas part are scaled in such a way that the volume integral matches the initial volume integral. The last step is to switch the sign of the smeared source terms in the liquid part to account for the mass loss. We end up with two source term distributions with different signs on the two sides of the interface where the volume integral of the absolute value matches the initial source term distribution. With this operation, we subtract mass from the liquid side and add it on the gas side. Hence, no source terms are needed for the VoF equation because changes in the liquid content are handled by the pressure equation.

The advantage of this approach is that it is easy to implement and that it works for arbitrary cell shapes. The disadvantage is that the smearing may cause nonphysical removal of liquid and requires a fine resolution near the interface.

\subsubsection{HardtWondraGas}

This model is identical to the Hardt and Wondra model with the exception that the liquid loss is accounted for in the VoF equation. Hence, there are no source terms in the presure Poisson equation in the liquid. Instead, a sharp source is used in the VoF equation. 

With this approach velocities caused by the source terms in the liquid are no longer present. On the other hand, the shrinkage in a thin liquid film can be represented more accurately by the sharp source term which guarantees the removal or addition of liquid in the correct location. An additional benefit is that in some scenarios the stability of the solver can be improved since inaccuracies in the mass conservation can lead to the creation of cells with a tiny amount of liquid ($\alpha < 1e-6$). In the case of the geometric VoF scheme, a small interface segment would be found in that cell which may cause a significant amount of phase change in that region. With the Hardt and Wondra model \cite{Hardt.2008} these tiny liquid volumes cannot evaporate and may accumulate during the simulation.

\subsection{Surface force module}

One challenge in the simulation of multiphase flows is the reduction of parasitic currents. These are artificial velocities which are induced around the interface by an imperfect discretization of the pressure jump condition. This effect is well known for surface tension driven flow but can also arise from external accelerations \cite{Wroniszewski.2014} \cite{RoenbyJohanandBredmoseHenrikandJasakHrvoje.2019}. 

\begin{figure}
	\centering
	\includegraphics[width=0.7\linewidth]{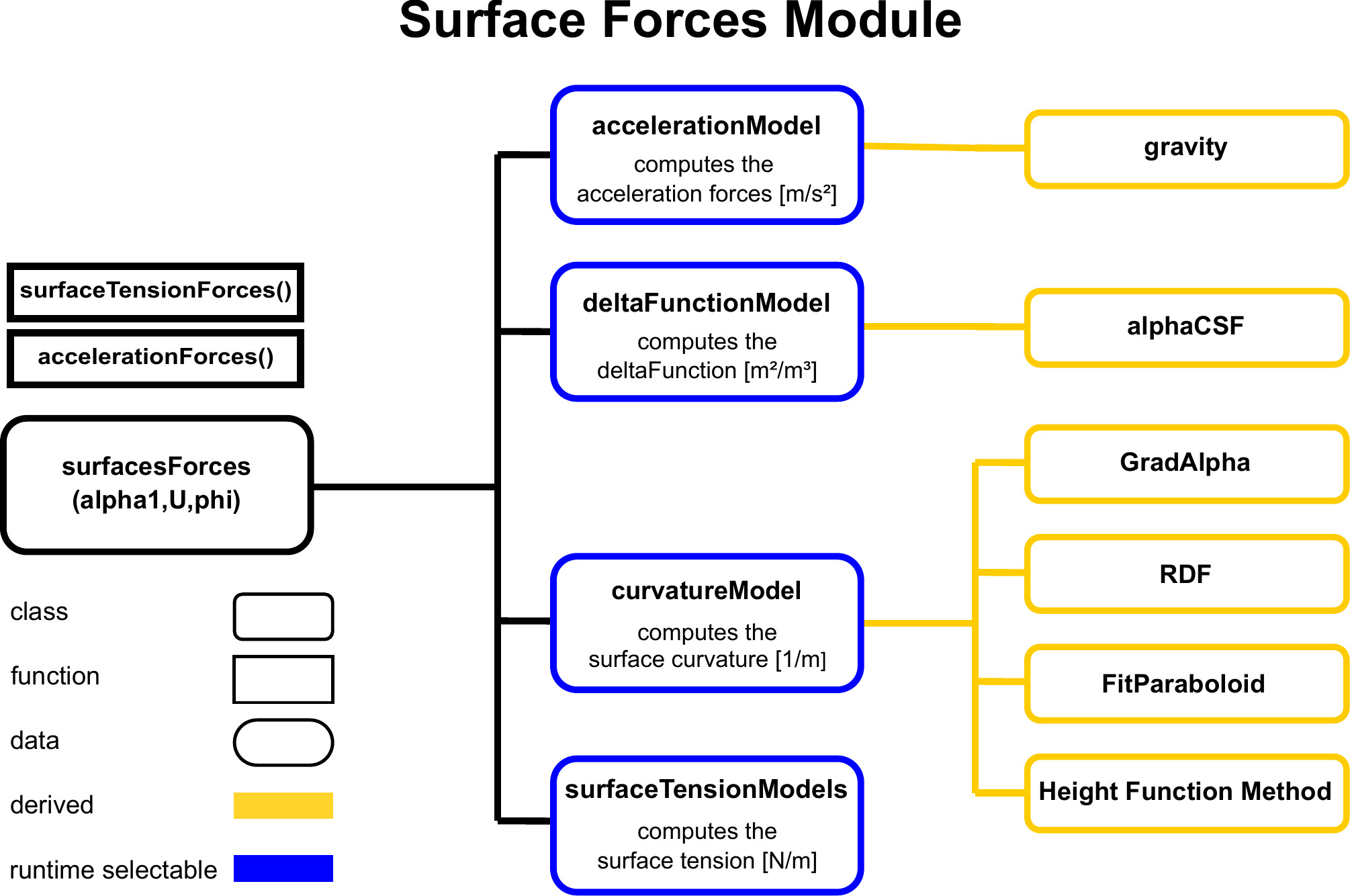}
	\caption{Schematic overview of the surface forces module}
	\label{fig:schematic_SurfaceForcesModel}
\end{figure}

Both the surface tension and acceleration forces induce a pressure jump over the interface. In the case of surface tension, this is the expected behaviour. As described earlier, the OpenFOAM interfacial flow solvers, interFoam, interIsoFoam and compressibleInterFoam employ a pressure where the hydrostatic potential is subtracted, $p_\textrm{rgh} = p - \rho \vek g\cdot \vek x$. This formulation reduces the spurious currents \cite{Wroniszewski.2014} and simplifies the definition of the hydrostatic pressure boundary condition \cite{Popinet.2017}. The acceleration force can be written in the form,
\begin{equation}
	\vek F_{a,f} = (\rho^{l}-\rho^{v})(\vek a \cdot \vek x) \nif \delta_s,
	\label{eq:generalFa}
\end{equation}
where $\mathbf{a}$ is the acceleration vector. The surface tension takes a similar form,
\begin{equation}
	\vek F_\textrm{st,f} = \sigma \kappa \nif \delta_s,
	\label{eq:Fst}
\end{equation}
with the surface tension $\sigma$ and the curvature $\kappa$. Both forces share the same mathematical background which is explained in Popinet \cite{Popinet.2017} and Ghidaglia \cite{Ghidaglia.2016} in more detail.  Both the acceleration force and surface tension force must be calculated on the face centre to achieve a well-balanced formulation with the pressure equation.
\sloppy
This formulation forms the basis of the surface forces module and is depicted in Fig. \ref{fig:schematic_SurfaceForcesModel}. It shows an overview of the implemented framework with runtime selectable classes and the currently available models which will be described in more detail below. The class, \texttt{surfaceForces},  handles the interface to the solver as shown above. The \texttt{accelerationModel} reflects the factor $\mathbf{a \cdot x}$ , the \texttt{deltaFunctionModel} the factor $\nif \delta_s$, the \texttt{curvatureModel} the factor $\kappa$ and \texttt{surfaceTensionModels} the factor $\sigma$. This design allows body force implementations based on other interface representations than VoF, e.g. LS or front-tracking, to be integrated.
Marangoni convection can be modelled with the \texttt{surfaceTensionModels} as it allows for the definition of temperature depended surface tension, $\sigma$. However, throughout this paper we assume that the surface tension is constant.
\fussy
\subsubsection{Delta function model}

The surface forces act on a microscopic region at the interface that is typically significantly smaller than the cell size. But the discretization of the function in the finite volume framework requires a region of at least one cell size to convert it into a volumetric force. There are multiple approaches found in literature \cite{Popinet.2017} to discretize $\nif \delta_s$. Currently, only the Continuous Surface Force (CSF) method proposed by Brackbill et al. \cite{BrackbillJeremiahU.DouglasB.KotheandCharlesZemach.1992} is available in our framework:
\begin{equation}
	\nif \delta_s = \nabla \alpha
\label{eq:deltaBrackbill}
\end{equation}

\subsubsection{Curvature model}

The prediction of the curvature is essential for an accurate simulation of the surface tension driven flow. If the exact curvature is known, the parasitic currents will drop to zero as the well-balanced formulation of the force is implemented in the proposed library and in the standard solver in OpenFOAM. The computation of the curvature is extremely challenging and numerous models have been proposed to tackle that issue. This library provides multiple implementations of curvature models that can be selected in the solver. 

\paragraph{gradAlpha}

This model is identical to the standard OpenFOAM formulation in the multiphase solvers. The model is based on the paper of Brackbill et al. \cite{BrackbillJeremiahU.DouglasB.KotheandCharlesZemach.1992} and computes the curvature directly from the VoF field, $\alpha$. 
The computation consists of two steps: 1) the computation of the normal with the default gradient operator:
\begin{equation}
 \hat{\mathbf{n}}_\alpha = \frac{\nabla \alpha}{|\nabla \alpha|}
 \label{eq:normalAlpha}
\end{equation}
2) Subsequently, the normals are interpolated to the faces, normalized and the curvature is computed as the divergence of the normals on the faces:
\begin{equation}
\kappa = \nabla \cdot \hat{\mathbf{n}}_\alpha
\label{eq:curvDivNormal}
\end{equation}
The contact angle in this implementation is treated by setting the normal on the boundary face to the prescribed contact angle. The implementation of this model is straightforward and is probably the most frequently used model for curvature computation in combination with VoF based multiphase solvers.

The accuracy of the model can be increased drastically by using a least square based gradient method \texttt{pointCellLeastSquares}.

\paragraph{Reconstructed distance function (RDF)}

The Reconstructed Distance Function model (RDF) is based on an implementation of the model proposed by Cummins et al. \cite{Cummins.2005}. The RDF model shares a lot of similarities with coupled LS-VoF models. The most significant difference is that in the RDF model the signed distance function, $\psi$, is not found by solving a PDE as in the LS method, but is instead constructed geometrically based on the volume fraction field. 

The first step of the algorithm is the reconstruction of the signed distance function, $\psi$, in a narrow band around the interface. The value of $\psi$ in every cell centre of the narrow band is computed as
\begin{equation}
    \psi_{cc} = \nif \cdot (\vek x_{cc} - \vek x_s).
\end{equation}
After that we calculate the gradient of the RDF:
\begin{equation}
\hat{\mathbf{n}}_\psi = \frac{\nabla \psi}{|\nabla \psi|}
\label{eq:normalRDF}
\end{equation}
By definition, the gradient of a signed distance function has length equal to one but due to discretization errors this is not necessarily the case after the numerical calculation. Therefore, we normalize the calulated vector by dividing by its length. 

The curvature is then computed by interpolating the normal, $\hat{\mathbf n}_\psi$, from cell centres to faces and applying the Gauss-Green gradient method,
\begin{equation}
    \kappa = \nabla \cdot \hat{\mathbf n}_\psi \rightarrow \kappa_{cc} \approx \frac1{V_{cc}} \sum_f \hat{\mathbf n}_{\psi,f}\cdot \vek S_f,
    \label{eq:curvNormal}
\end{equation}
where $V_{cc}$ is the cell volume, the sum is over all the cell's faces and $\vek S_f$ is the face area vector pointing out of the cell. This method is accurate on structured grids but causes inaccuracies on unstructured grids due to interpolation errors. These can be reduced by computing the curvature as
\begin{equation}
    \kappa = tr( \nabla \hat{\mathbf n}_\psi).
    \label{eq:curvTrNormal}
\end{equation}
While the two formulations are mathematically identical, the latter allows the usage of the numerical least square gradient method which is more accurate on unstructured grids because the Gauss-Green gradient method is only zeroth order accurate on unstructured grids \cite{syrakos2017}. After the computation of Eqn.~\ref{eq:curvTrNormal}, we have the curvature at the cell centres which would limit the scheme to first-order accuracy. The accuracy can be increased by interpolating the curvature to the interface centres using OpenFOAM's \texttt{cellPointInterpolation} class, which interpolates cell centred data to any point in the computational domain.

To improve the accuracy of the normal calculation near boundary faces, they are included in the stencil. The value of the signed distance function, $\psi$ on the boundary is computed with the consideration of the extrapolated interface information described in section \ref{sec:VoF_module}. After the computation of the normal with the gradient operator, the boundary values of the normals are set to the prescribed contact angle as in \texttt{gradAlpha} model.

The accuracy of the method can be significantly influenced by the accuracy of the gradient operator specified in \texttt{fvSchemes} which makes OpenFOAM's \texttt{pointCellLeastSquares} the recommended choice.

\paragraph{fitParaboloid}
\label{seq:fitParaboloid}

This method estimates the curvature by fitting a local function to the neighbour interface centre provided by the interface reconstruction scheme. The local function is approximated by:
\begin{equation}
\begin{array}{ll}
f(x,z) = C_0 x + C_1 x^2 + z  & \textrm{2D} \\
f(x,y,z) = C_0 x + C_1 x^2 + C_2 y + C_3 y^2 + C_4 xy + z & \textrm{3D} 
\end{array}
\label{eq:fitPoly}
\end{equation}
This equation is solved with a least square minimization by rotating the coordinate system with the $z$-direction aligned with the interface normal. The resulting linear system can be efficiently solved with an LU factorization. With the so obtained coefficient, the derivatives of the above function can now be calculated analytically and the curvature computed as
\begin{equation}
\kappa = \left\{
\begin{array}{ll}
\frac{f_{xx}}{(1+f_x^2)^{3/2}} & \textrm{2D} \\ [10pt]
\frac{f_{xx}(1+f_x) + f_{yy}(1+f_y) - 2 f_x f_y f_{xy}}{(1+f_x^2+f_y^2)^{3/2}} & \textrm{3D} \\
\end{array}
\right. 
\label{eq:curv_exp}
\end{equation}
The handling of the prescribed boundary condition is straightforward and achieved by including the extrapolated centre value in the stencil (see section \ref{sec:VoF_module}).

\paragraph{Height Function Method}

\begin{figure}
	\centering
	\begin{minipage}{0.48\textwidth}
		\centering
		\includegraphics[width=0.9\textwidth]{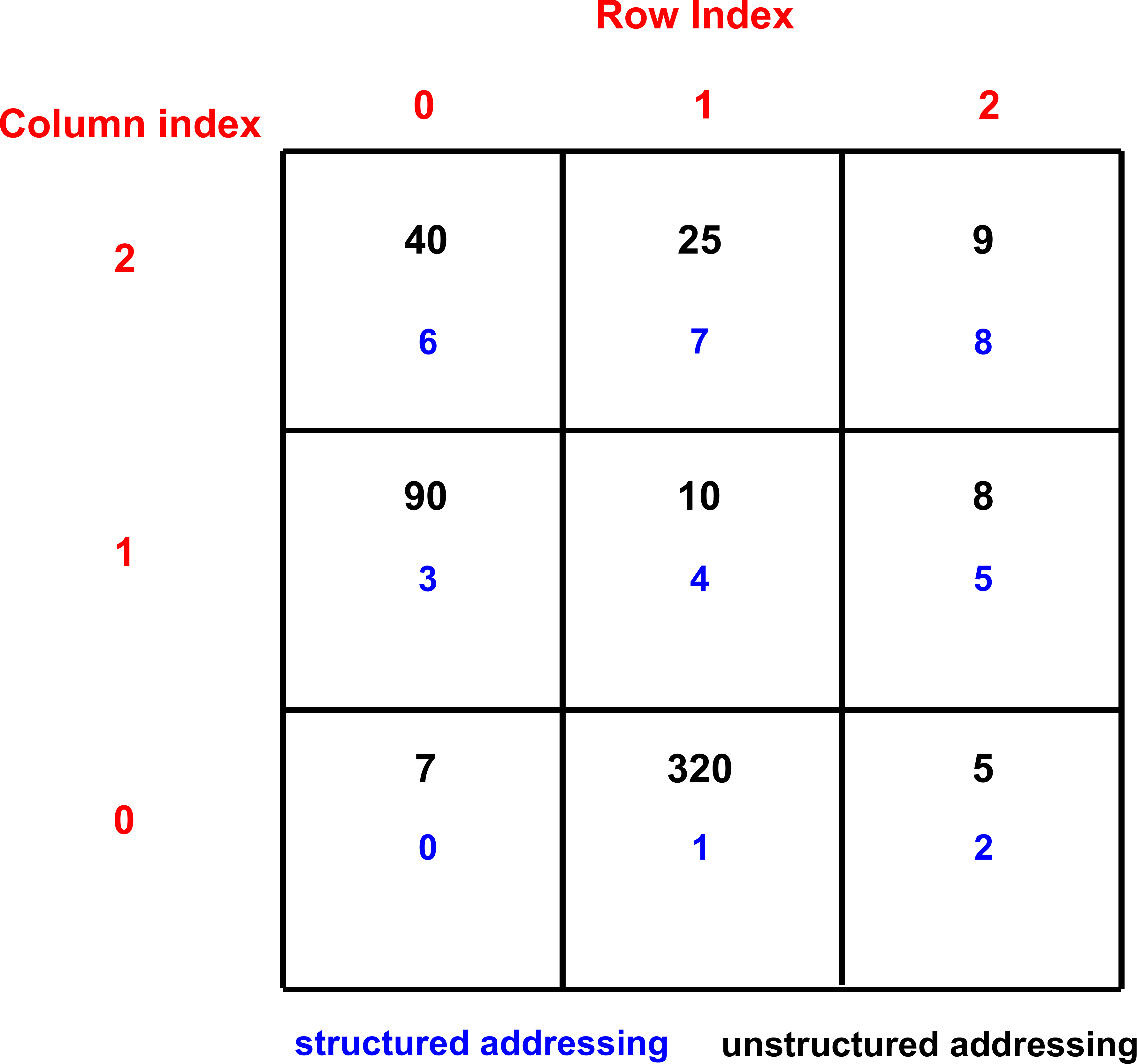} 
		\caption{Height function stencil.}
		\label{fig:HFStencil}
	\end{minipage}\hfill
	\begin{minipage}{0.48\textwidth}
		\centering
		\includegraphics[width=0.9\textwidth]{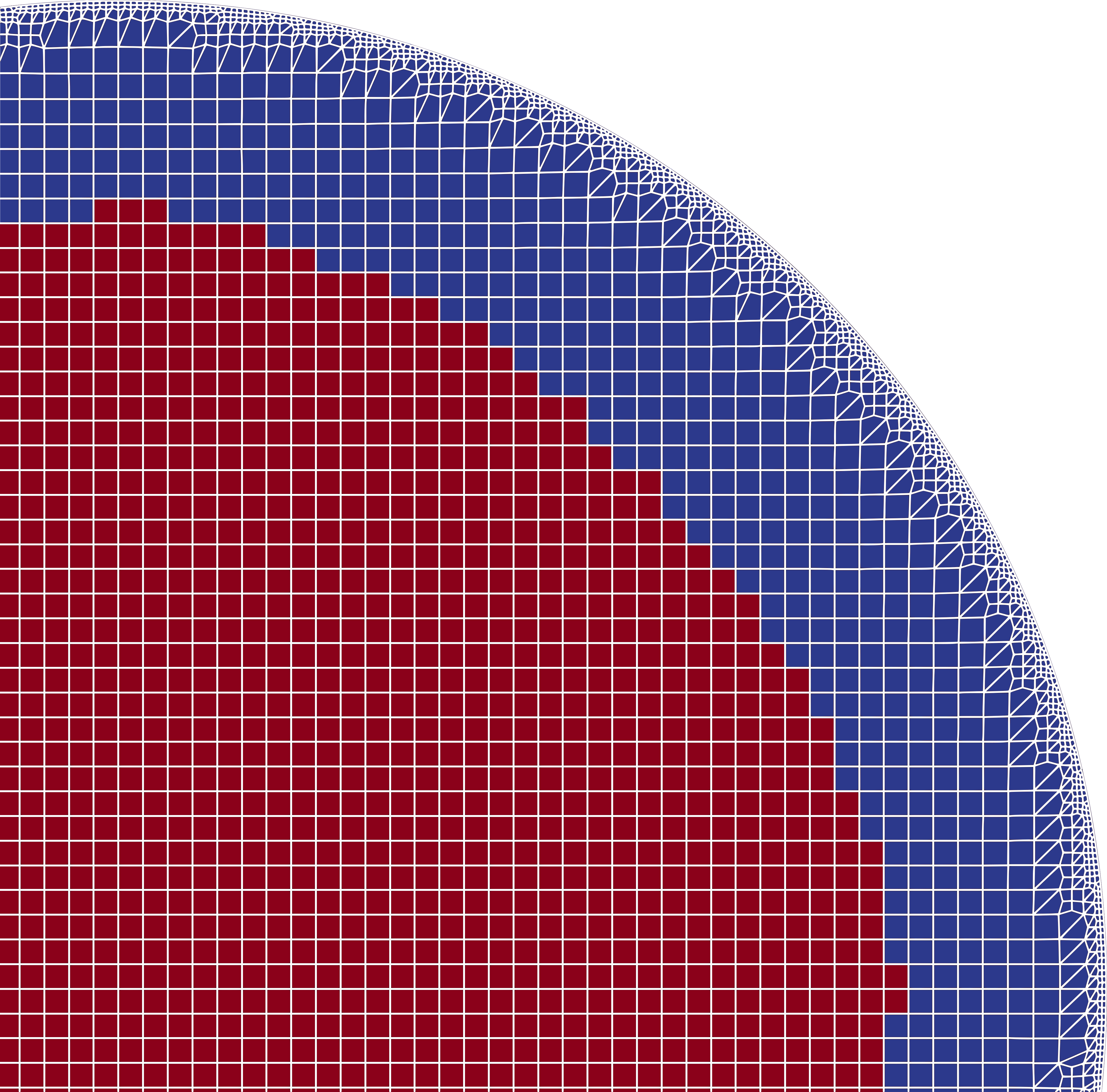} 
		\caption{Cubic cells are marked in red.}
		\label{fig:Cyl_mesh}
	\end{minipage}
\end{figure}

The Height Function Method (HFM) is a simple and second-order accurate method for the computation of curvature \cite{Popinet.2009}. The implementation is straightforward on structured grids and is successfully used in the basilisk flow solver \cite{basilisk2020}.
Attempts to implement this method on unstructured grids result in considerably more complex implementation but so far without achieving the second-order accuracy \cite{Owkes.2015} \cite{Evrard.2017}. The following variant of the height function method is only utilized in structured parts of the mesh where the mesh elements are cubes. The open source meshing tools cfMesh \cite{cfMesh2017} or snappyHexMesh \cite{snappyHexMesh2020} generate a hex mesh in the interior of the domain as depicted in Fig. \ref{fig:Cyl_mesh}. With this hybrid approach, complex geometries can be represented with a body fitted mesh and the accurate height function method can be utilized on the majority of the mesh. The general outline of the method is given in Algorithm \ref{alg:HFMethod}. 

{\begin{algorithm}
		\footnotesize
		\DontPrintSemicolon
		\SetAlgoNoLine
		Classify cuboid cells \\
		Calculate interfaceCells: A list of all the interface cell indices, i.e. $i\in$ interfaceCells if $\epsilon < \alpha_i < 1-\epsilon$.\\
		\For {celli in interfaceCells}
		{
			\eIf{isCuboid}
			{
				vector n = surfaceNormal[celli].normalize() \\
				\For {dir in sortAbsoluteComponents(n)}
				{
					\tcp{function computeHeight is described in Algorithm \ref{alg:computeHeight}}
					curv, foundHeight = computeHeight(dir) \\
					curvature[celli] = curv \\
					\If{foundHeight}
					{
						break \\
					}
				}
				\If{not foundHeight}
				{
					curvature[celli] = fitParaboloid() \\
				}
			}
			{
				curvature[celli] = fitParaboloid() \\	
			}
			
		}
		\caption{Outline of the height function method}
		\label{alg:HFMethod}
\end{algorithm}}

First, we have to classify the cubic cells in the grid. A cell is defined as cubic, if 1) it has six faces and 2) all vectors from the cell centre to the neighbour cell centres form an orthogonal base (with an angle tolerance of 0.001 degrees. An example of such a classification is depicted in Fig. \ref{fig:Cyl_mesh} where cells marked red are considered cubes. If an interface cell is not a cube, or the height function method fails, a paraboloid is fitted in the neighbouring interface centres as described in section \ref{seq:fitParaboloid}.

\begin{figure}
	\centering
	\includegraphics[width=0.6\linewidth]{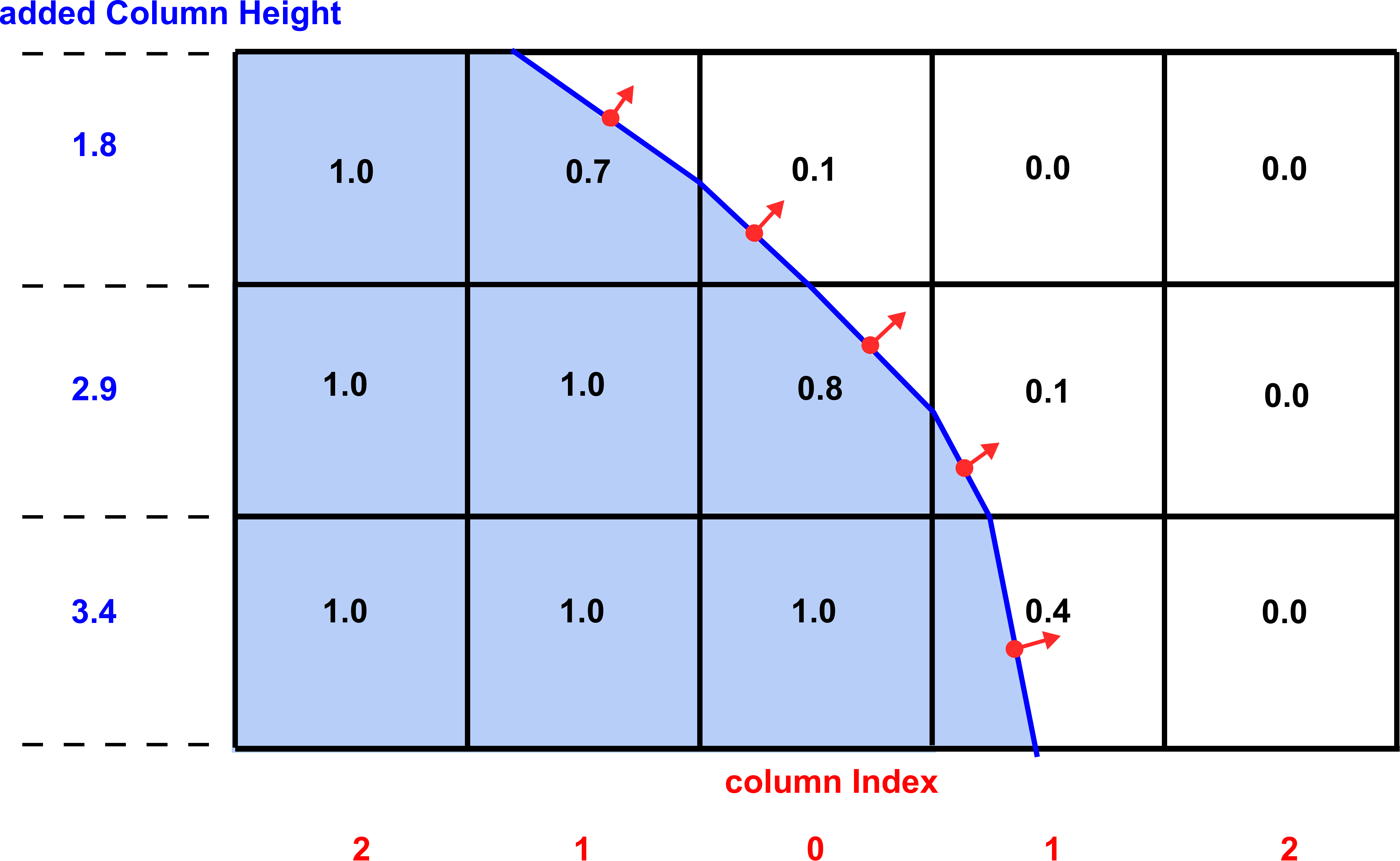}
	\caption{Volume of fluid field with the computed column heights}
	\label{fig:HFMethod}
\end{figure}

The height function method estimates the curvature directly from the VoF field similar to the method proposed by Brackbill \cite{BrackbillJeremiahU.DouglasB.KotheandCharlesZemach.1992}. However, it does not compute the curvature with the gradient of the VoF field but instead with the column height. These columns represent the interface in the form $H(x,y,f(x,y))$ and the derivatives of this function $H$ are used to compute the curvature given in Eqn.~\ref{eq:curv_exp}. The derivatives are computed with a central second order accurate finite difference operator. The height function is accumulated volume fraction field in the coordinate directions, $x, y$ or $z$, of the structured grid. Fig. \ref{fig:HFMethod} shows computed column heights for a structured grid in $x$-direction. The computation of the heights for a given cell is considered successful if  all columns at a given column index only contain liquid on one side and only gas on the other side. This is the case in Fig. \ref{fig:HFMethod} if we look two cells to the right and left. So the average of the column to the left of the surface would be one and the average of the surface to the right of the interface would be zero.

Algorithm \ref{alg:computeHeight} describes the procedure of adding the heights in the given direction.  The basic idea is to compute the height for the first column index (see Fig. \ref{fig:HFMethod}) and then to advance in positive and negative direction and add the height in that location.  After the accumulation of the height values, second-order accurate finite difference operators are used to compute the heights.

{\begin{algorithm}
		\footnotesize
		\SetAlgoNoLine
		\SetKwProg{Def}{def}{:}{}
		\Def{computeHeight(direction)}
		{
			\tcp{construct 2D direction-aligned stencil }
			twoDimFDStencil cols(celli,dir) \\
			\tcp{compute column height for col index 0}
			cols.addColumnHeight(celli) \\				
			label $dir_{pos}$, $dir_{neg}$ = cols.nextCellsInDirection() \\
			\tcp{advance the col index in pos and neg direction Fig.\ref{fig:HFMethod}}
			avgHeightValPos, avgHeightValNeg = 0.5 \\
			\For{iter in [1 ... 7]}
			{
				\tcp{is column empty avgHeight = 0 , full avgHeight = 1}
				\If{foundHeight(avgHeightValPos)}
				{
					avgHeightValPos = cols.addColumnHeight($dir_{pos}$) \\
					label $dir_{pos}$ = cols.nextCellsInDirection(pos=True) \\
				}
				
				\If{foundHeight(avgHeightValNeg)}
				{
					avgHeightValNeg = cols.addColumnHeight($dir_{neg}$) \\
					label $dir_{neg}$ =  cols.nextCellsInDirection(pos=False) \\
				}
			}
			\tcp{full and empty col found?}
			bool foundHeight = foundHeight(avgHeightValPos,avgHeightValNeg) \\
			scalar curv = cols.calcCurvature() \\
			return curv, foundHeight;
		}
		\caption{ComputeHeights}
		\label{alg:computeHeight}
\end{algorithm}}

To advance in the given direction, the unstructured cell-point-cell stencil needs to be sorted to mimic structured stencil addressing as depicted in Fig. \ref{fig:HFStencil}. The position in the stencil of 27 (in 3D) or 9 (in 2D) cells is calculated with structured coordinate directions $i, j$ and $k$ as position $= i + 3j + 9k$. With this addressing, the next cell label in the unstructured grid for the given positive and negative direction can easily be computed.

In parallelised cases, where the next cell in the current direction is on a neighbour processor, this neighbour processor will continue to accumulate heights. For this task the neighbour processor needs the direction and status of the iteration as well as the cell index for which the height functions need to be computed. With this information, the neighbour processor can perform the height accumulation and send the results back to the original processor. 

\subsection{Acceleration model}

It is well-documented in literature that for surface tension dominated flows spurious currents can arise at the fluid interface due to inaccurate curvature estimation. Perhaps less well--known are the spurious currents arising from numerical errors introduced in the discretisation of the gravity term (second term on the right hand side of Eqn.~\ref{eq:NSE}). There is a need for experimenting with different solutions to this problem. The framework offers the possibility to implement new gravity or acceleration models. Currently, only a single \texttt{gravity} model is available which is identical to the standard OpenFOAM implementation, i.e., 
\begin{equation}\label{eq:Fa}
		\vek F_{a,f} = (\mathbf g \cdot \mathbf x_f) \hat{\vek n}_f\cdot\nabla \rho,
\end{equation}
where $\mathbf x_f$ is the face centre and $\hat{\vek n}_f$ the unit normal to the face.

\section{Validation}

In the following, several benchmark test cases for the validation of the surface tension and phase change model are presented and compared to analytical solutions. 

The framework allows easy implementation of new models and simplifies the testing by providing established numerical benchmarks. To simplify the parameter studies needed to verify the implementation and compare results with different models, we use the open-source library caseFoam \cite{caseFoam2019}. It provides an easy way of generating series of simulation test cases where various parameters are  scanned within a specified range. The post processing data of the parameter studies can then easily be analysed and compared.

\subsection{Phase change}

The easiest and most accurate way to test an implementation is to compare its results with known analytical solutions. For the phase change model, three analytical solutions are widely used: The Stefan problem, the sucking interface problem and the Scriven problem. The different models can be selected by changing the entries in the \texttt{phaseChangeProperties} dictionary:
\begin{minipage}{\linewidth}
\begin{footnotesize}
\begin{verbatim}
// options: selectedGradExplicit implicitGrad Schrage
energySourceTermModel implicitGrad; 

implicitGradCoeffs
{
}

// options: hardtWondra hardtWondraGasPhase
massSourceTermModel hardtWondra;
hardtWondraCoeffs
{
} 

satProperties
{
    singleComponentSatProp function;
    Tmin 100;
    Tmax 500;
    pSat constant 1e5;
    TSat constant 373.15;
    L constant 2.26e6;
}
\end{verbatim}
\end{footnotesize}
\end{minipage}

\subsubsection{Stefan problem}

\begin{figure}
    \begin{subfigure}{0.5\textwidth}
        \includegraphics[width=\linewidth]{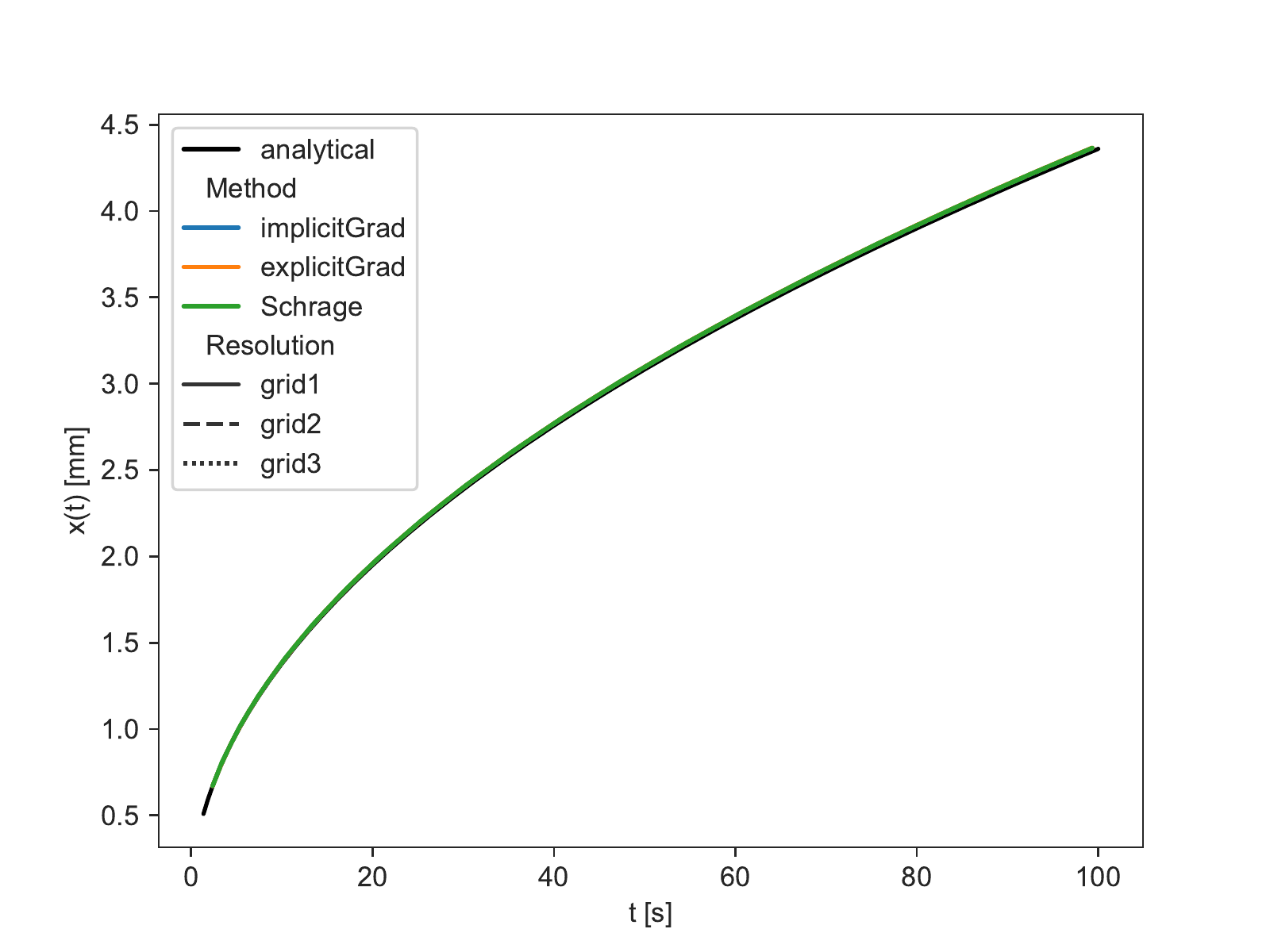}
        \caption{Geometric VoF (isoAdector).} \label{fig:Stefan_plicRDF}
        \end{subfigure}
        \hspace*{\fill} 
    \begin{subfigure}{0.5\textwidth}
        \includegraphics[width=\linewidth]{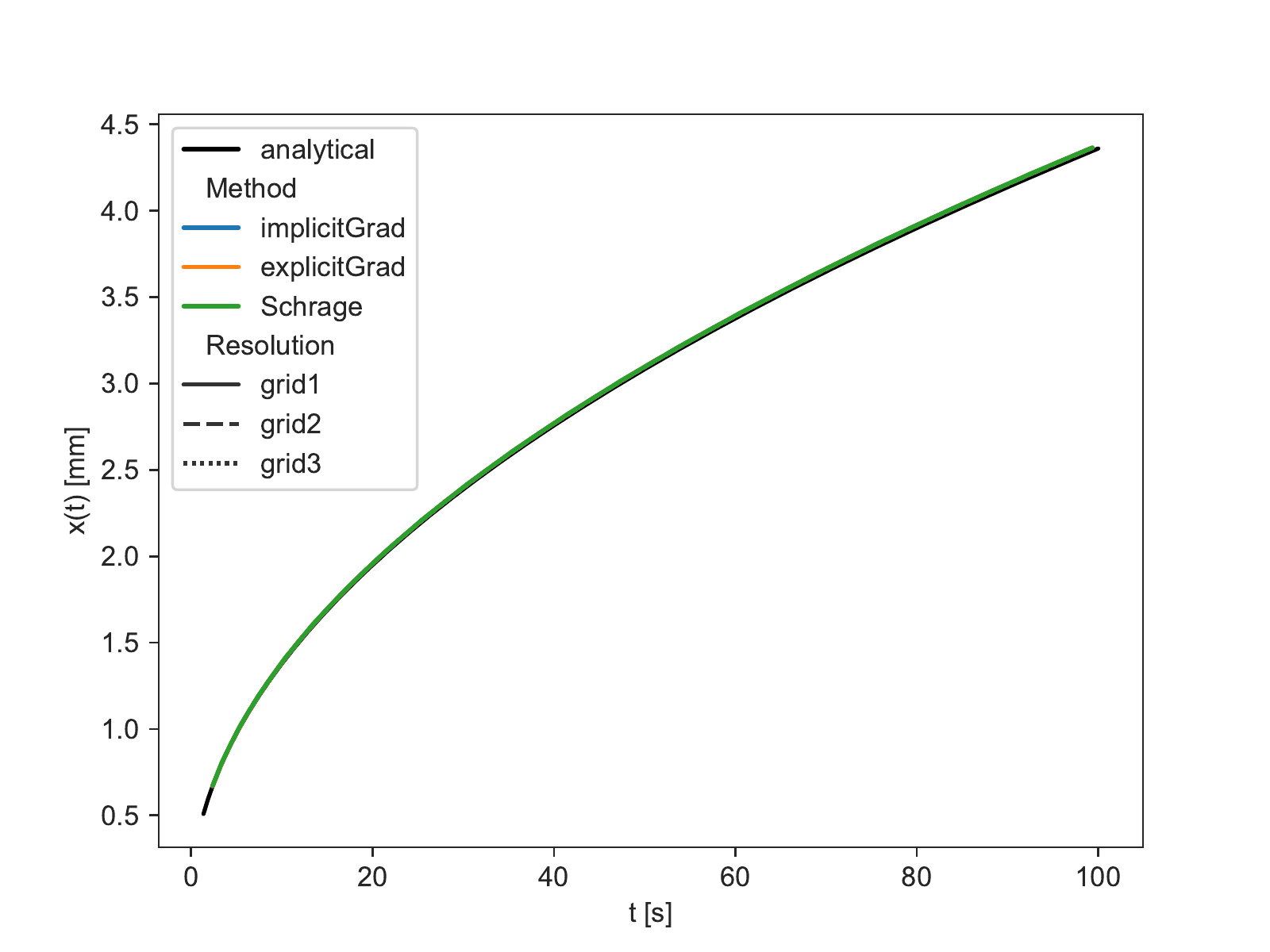}
        \caption{Colour function VoF (MULES)} \label{fig:Stefan_isoSurface}
    \end{subfigure}
    \caption{Stefan problem: Comparison of the interface position} \label{fig:StefanProblem}
\end{figure}

This one-dimensional validation test case describes the interface motion away from a superheated wall and is one of the most frequently used validation test cases \cite{Sato.2013} \cite{Kunkelmann.2011} \cite{Batzdorf.2015} \cite{Hardt.2008}. A gas column separates the superheated wall and the liquid. Due to energy transfer from the wall through the gas phase to the liquid interface, evaporation at the interface occurs, causing an interface displacement away from the wall. The analytical interface motion is given by
\begin{equation}
x(t) =  2 \beta \sqrt{a_{v} t}.
\label{eq:posStefan} 
\end{equation}
where $a_v$ is the thermal diffusivity, and $\beta$ is the solution to the transcendental equation (see \cite{Sato.2013})
\begin{equation}
\beta \cdot \mathrm{exp}(\beta^{2}) \cdot \mathrm{erf}({\beta}) =  \frac{c_p^v (T_{\mathrm{Wall}}-T_{\mathrm{Sat}})}{\sqrt{\pi}L}.
\label{eq:beta}
\end{equation}
In our test case the wall is superheated with $T_{\mathrm{Wall}} = T_{\mathrm{Sat}} + 5 K$ and the thermodynamic properties of the fluids are given in Tab. \ref{tab:thermalPropScriven}. The grid is one-dimensional and has a length of 10 mm with a resolution of 50 (Grid1), 100 (Grid2) or 200 cells (Grid3). The solutions of the simulations are shown in Fig. \ref{fig:Stefan_plicRDF} and \ref{fig:Stefan_isoSurface}. All models deliver accurate results for all grid sizes and advection schemes.  

\begin{table}[htbp]
	\centering
	\caption{Thermal properties of the mesh regions}
	\label{tab:thermalPropScriven}
		\begin{tabular}{c|ccccc}
		  & $ R \: \mathrm{[J/(kg K)]} $ & $\rho \: \mathrm{[kg/m^{3}]} $ & $c_{p} \: \mathrm{[J/(kg K)]} $ & $\lambda \: \mathrm{[W/(m K)]} $ & $ L \: \mathrm{[kJ/kg]} $ \\ \hline\hline
			vapor & 461.4 & 0.581 & 2030  & 0.025 & 2260 \\
			liquid & - & 958.4 & 4216 & 0.671 & 2260 \\
		\end{tabular}
\end{table}

\subsubsection{Sucking interface}

Another frequently encountered and slightly more complex one-dimensional test case is the sucking interface problem introduced by Welch and Wilson \cite{Welch.2000} which became one of the standard test cases \cite{Kunkelmann.2011} \cite{Sato.2013}. The model describes the interface movement for a one-dimensional superheated column. As in the Stefan problem, gas is located between a wall and the liquid but here the liquid is superheated and both gas and wall are on saturation temperature. The superheated liquid evaporates at the interface, creating volume pushing the liquid away from the wall. 

\begin{figure}
    \begin{subfigure}{0.5\textwidth}
        \includegraphics[width=\linewidth]{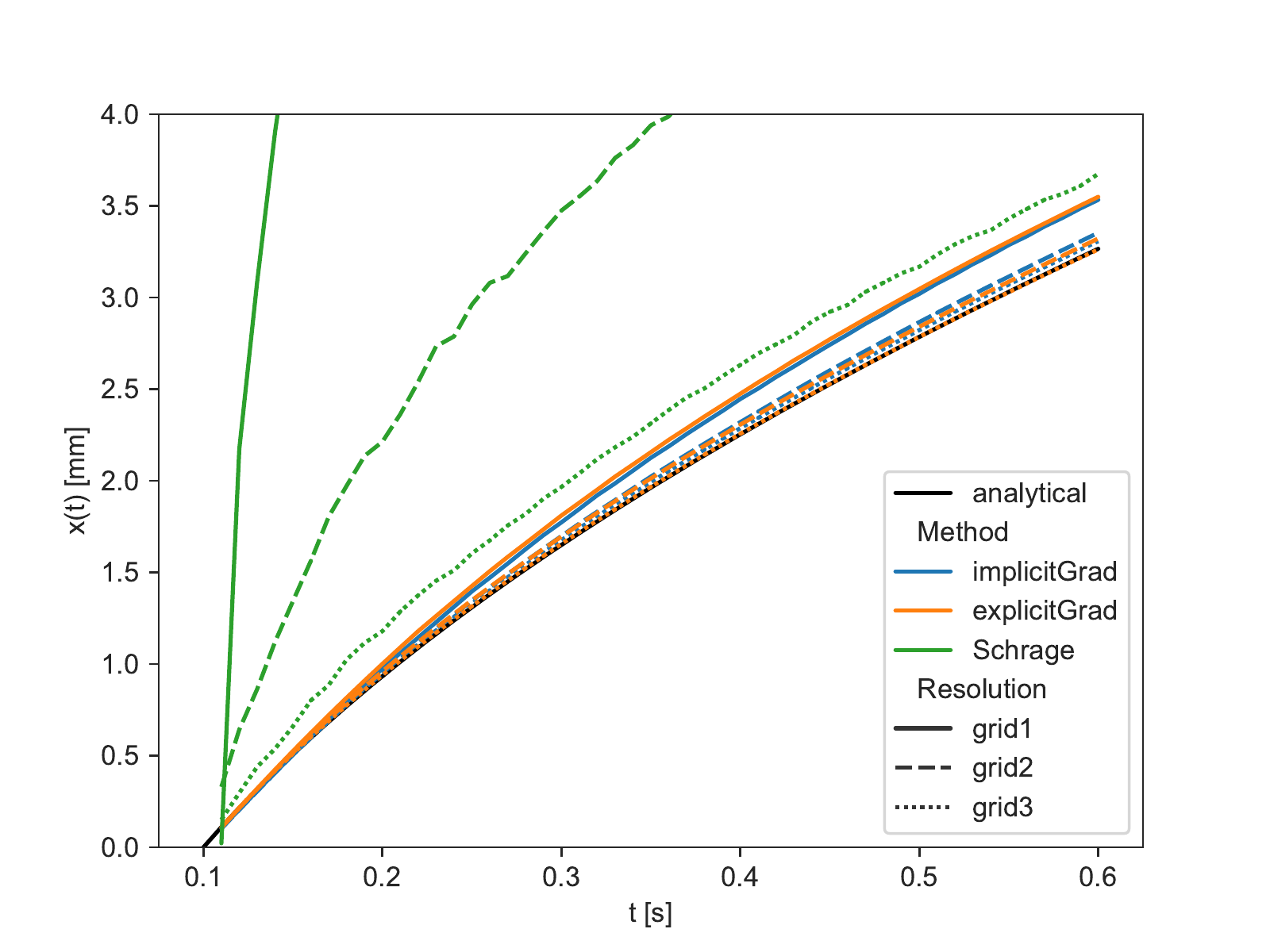}
        \caption{Geometric VoF (isoAdector)} \label{fig:suckingInterface_plicRDF}
        \end{subfigure}
        \hspace*{\fill} 
    \begin{subfigure}{0.5\textwidth}
        \includegraphics[width=\linewidth]{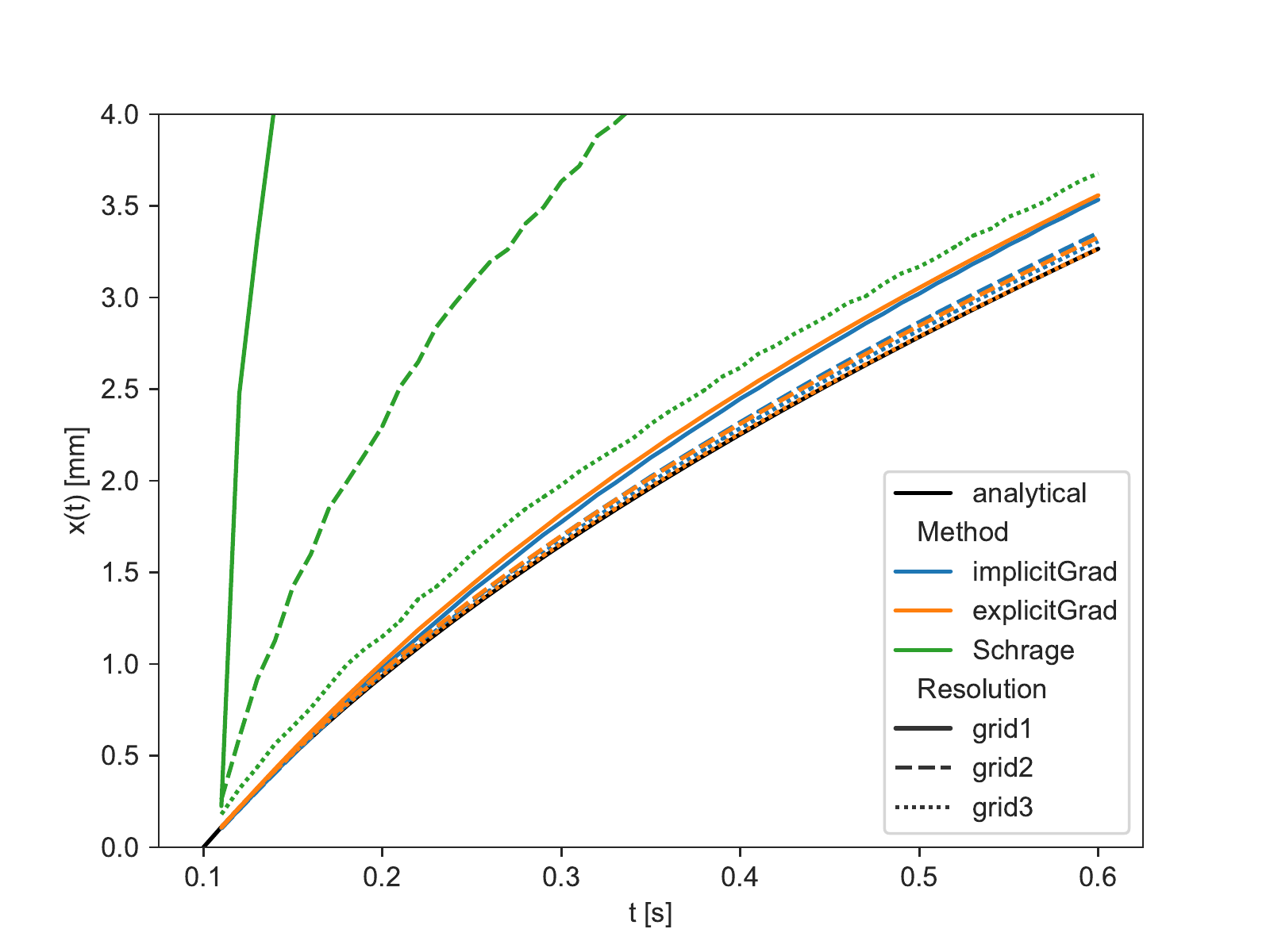}
        \caption{Colour function VoF (MULES)} \label{fig:suckingInterface_isoSurface}
    \end{subfigure}
    \caption{Sucking interface: Comparison of the interface position} \label{fig:SuckingInterface}
\end{figure}

In our case setup, the values of the analytical solution are given with the thermophysical properties of Tab. \ref{tab:thermalPropScriven} and with a superheated fluid temperature of 5 Kelvin. As in the previous simulation, the length of the domain is 10 mm but we now use grids with cell counts 100, 200 and 400. 
Fig. \ref{fig:suckingInterface_plicRDF} and \ref{fig:suckingInterface_isoSurface} show the parameter study for the three grid sizes and the available models. The influence of the advection scheme is observed to be neglectable, whereas a strong dependency of the choice of phase change treatment can be observed. The most accurate is the \texttt{implicitGrad} model followed by the \texttt{explictGrad} model whereas the \texttt{Schrage} model  is significantly less accurate.

\subsubsection{Bubble in superheated liquid}

The last phase change benchmark test case included here is the bubble in a superheated liquid. An analytical solution for this test case was provided by Scriven \cite{Scriven.1959} which was adopted as a benchmark test case  \cite{Kunkelmann.2011} \cite{Sato.2013} \cite{Batzdorf.2015} for phase change models. The analytical evolution of the radius is similar to the Stefan Problem and is given by
\begin{equation}
x(t) =  2 \beta_{\mathrm S} \sqrt{\alpha_{\mathrm l} t},
\label{eq:posScriven} 
\end{equation}
where $\beta_{\mathrm S}$ is the solution of the transcendental equation, 

\begin{equation}
\frac{\rho^l c_p^l (T_{\infty} - T_{\mathrm{Sat}})}{\rho^v(L + (c_p^l - c_p^v)(T_{\infty} - T_{\mathrm{Sat}}))} =  2 \beta_{\mathrm S}^{2} \int_0^1 e^{-\beta_{\mathrm S}^{2} \left( (1-\xi)^{-2} -2 \left(1 - \frac{\rho^v}{\rho^l}\right)\xi -1 \right)}d\xi.
\end{equation}
The initial bubble radius, $R$, and radial temperature distribution must be specified. For $r \leq R$ we set the initial $T = T_{\textrm{Sat}}$. For $r> R$ we use 

\begin{equation}
   T = T_{\infty} -  2 \beta_{\mathrm S}^{2} \frac{\rho^v(L + (c_p^l-c_p^v)(T_\infty - T_{\mathrm{Sat}})}{\rho^lc_p^l} \\ 
    \times \int_{1-R/r}^1 e^{-\beta_{\mathrm S}^{2} \left( (1-\xi)^{-2} -2 \left(1-\frac{\rho^v}{\rho^l}\right)\xi -1 \right)}d\xi
\end{equation}

\begin{figure}
    \begin{subfigure}{0.5\textwidth}
        \includegraphics[width=\linewidth]{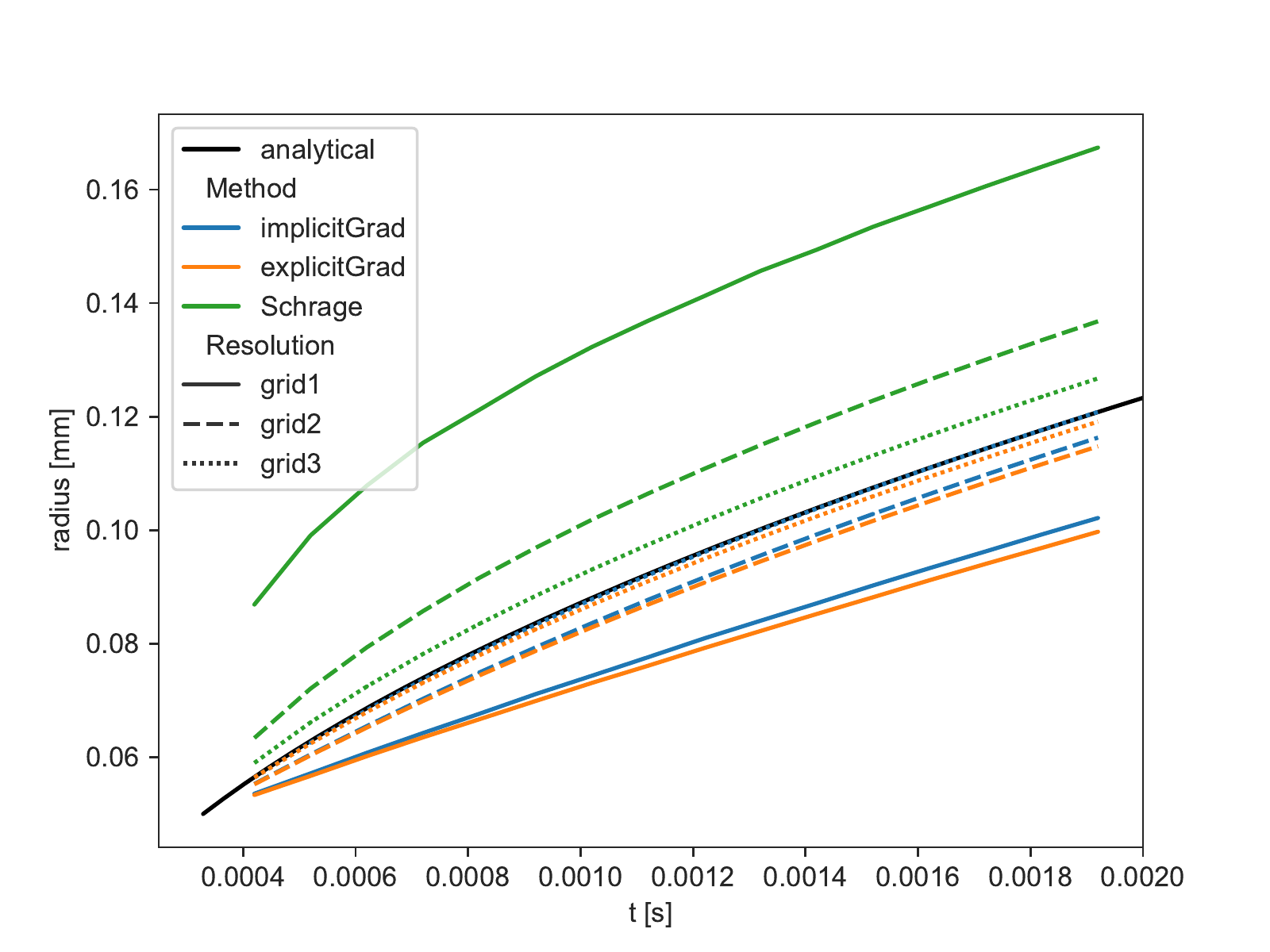}
        \caption{Geometric VoF (isoAdector)} \label{fig:Scriven_plicRDF}
        \end{subfigure}
        \hspace*{\fill} 
    \begin{subfigure}{0.5\textwidth}
        \includegraphics[width=\linewidth]{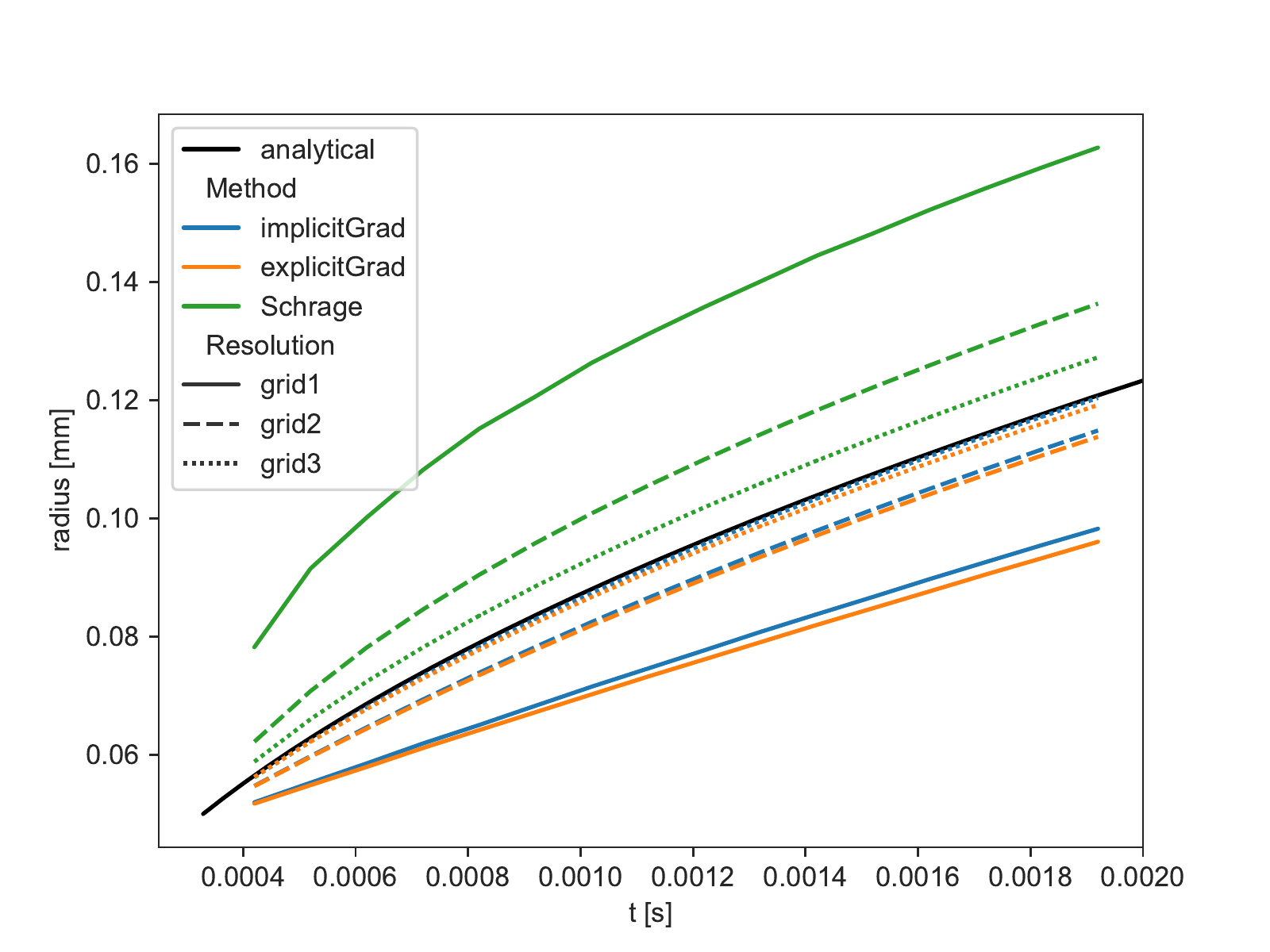}
        \caption{Colour function VoF (MULES)} \label{fig:Scriven_isoSurface}
    \end{subfigure}
    \caption{Bubble in superheated liquid: Comparison of the interface position} \label{fig:Scriven}
\end{figure}

Results are shown in Fig.~\ref{fig:Scriven}. As for the sucking interface problem, the gradient based schemes deliver the most accurate result followed by the Schrage results which behave similarly to the sucking interface. In contrast to the previous two test cases, the combination of geometric VoF and phase change model is slightly more accurate compared to the MULES advection scheme. 

\subsection{Surface tension models}

\emergencystretch 4em%
For the validation of the surface tension models multiple test cases, ranging from static reconstruction test cases to moving interface cases with non-constant curvature, are available. As in the previous examples different curvature models can be selected by changing the entry in the \texttt{transportProperties} or \texttt{thermophysicalProperties} depending on the solver:

\noindent
\begin{minipage}{\linewidth}
\begin{footnotesize}

\begin{verbatim}
surfaceForces
{
    sigma 0.01; // surface tension in SI units
    // options: gradAlpha RDF heightFunction fitParaboloid
    curvatureModel gradAlpha;
    accelerationModel gravity;
    deltaFunctionModel alphaCSF;
}
\end{verbatim}
\end{footnotesize}
\end{minipage}

\fussy
The most challenging part in the simulation of surface tension is an accurate description of the pressure jump at the interface. The proposed models simulate the pressure jump by applying forces at the mesh face centres which are proportional to the curvature. OpenFOAM uses the well-balanced surface tension formulation \cite{Deshpande2012} , resulting in an accuracy of machine precision if the correct curvature is specified. Accordingly, the simplest way of getting an accurate surface force model is by computing the curvature accurately. However, this is non-trivial as the curvature is proportional to the second derivative leading to large errors in the curvature computation if small errors in the basis function are observable. We use two curvature error measures: The average error, 

\begin{equation*}
	E_\textrm{Curv}^1 = \frac{1}{\kappa_\textrm{exact}}\sum_i^{N_\textrm{ic}} \frac{|\kappa_i - \kappa_\textrm{exact}|}{N_\textrm{ic}},
\end{equation*}
and the maximum error,
\begin{equation*}
	E_\textrm{Curv}^\textrm{max} = \frac{1}{\kappa_\textrm{exact}} \max_i (|\kappa_i - \kappa_\textrm{exact}|),
\end{equation*}
where the sum and max are over all the $N_\textrm{ic}$ interface cells. As described above, four models for the curvature computation are available with the HFM only working on structured grids.

\begin{figure}
    \begin{subfigure}{0.5\textwidth}
        \includegraphics[width=\linewidth]{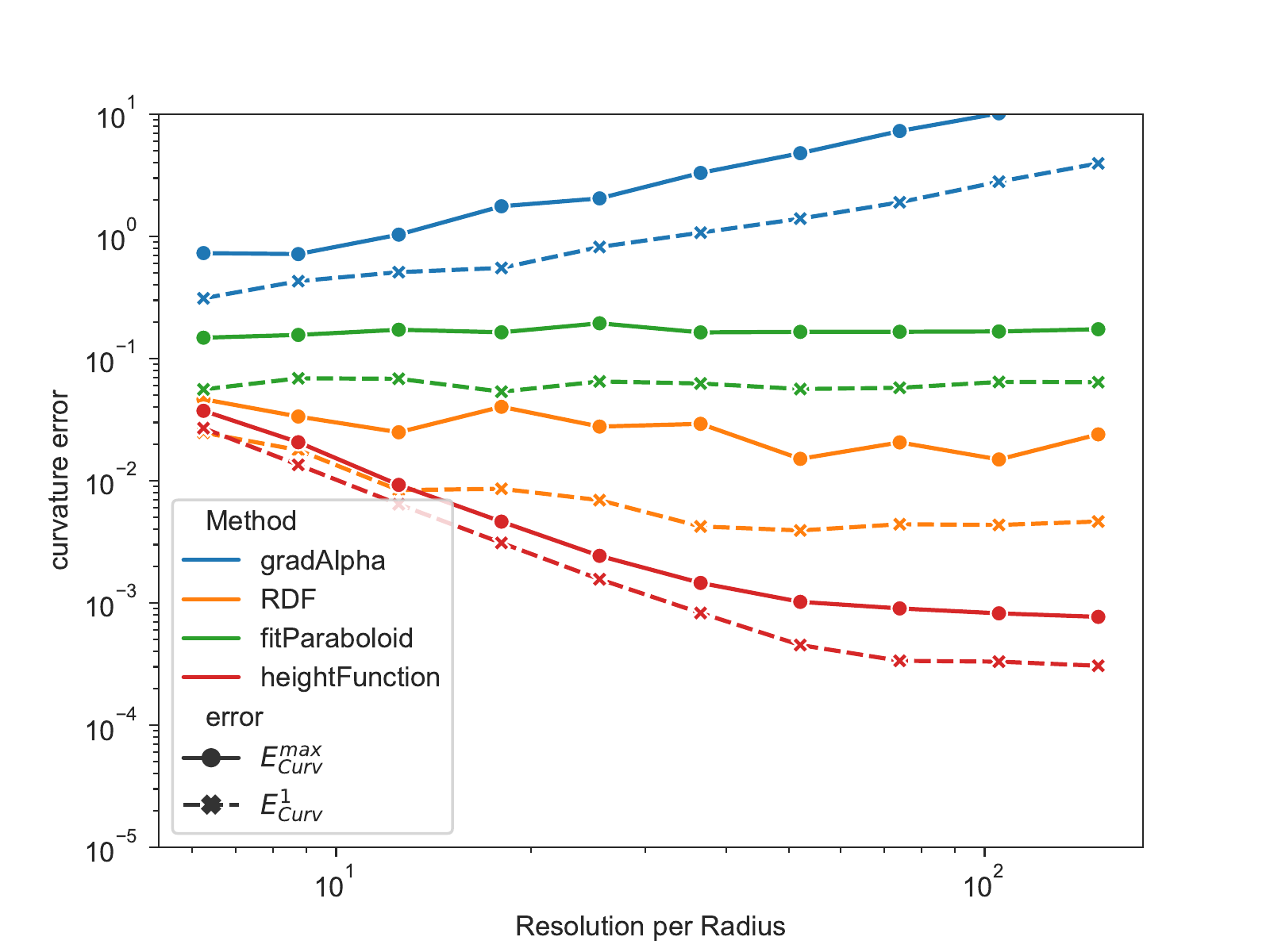}
        \caption{Hexahedral grid} \label{fig:curv2d_hex}
        \end{subfigure}
        \hspace*{\fill} 
    \begin{subfigure}{0.5\textwidth}
        \includegraphics[width=\linewidth]{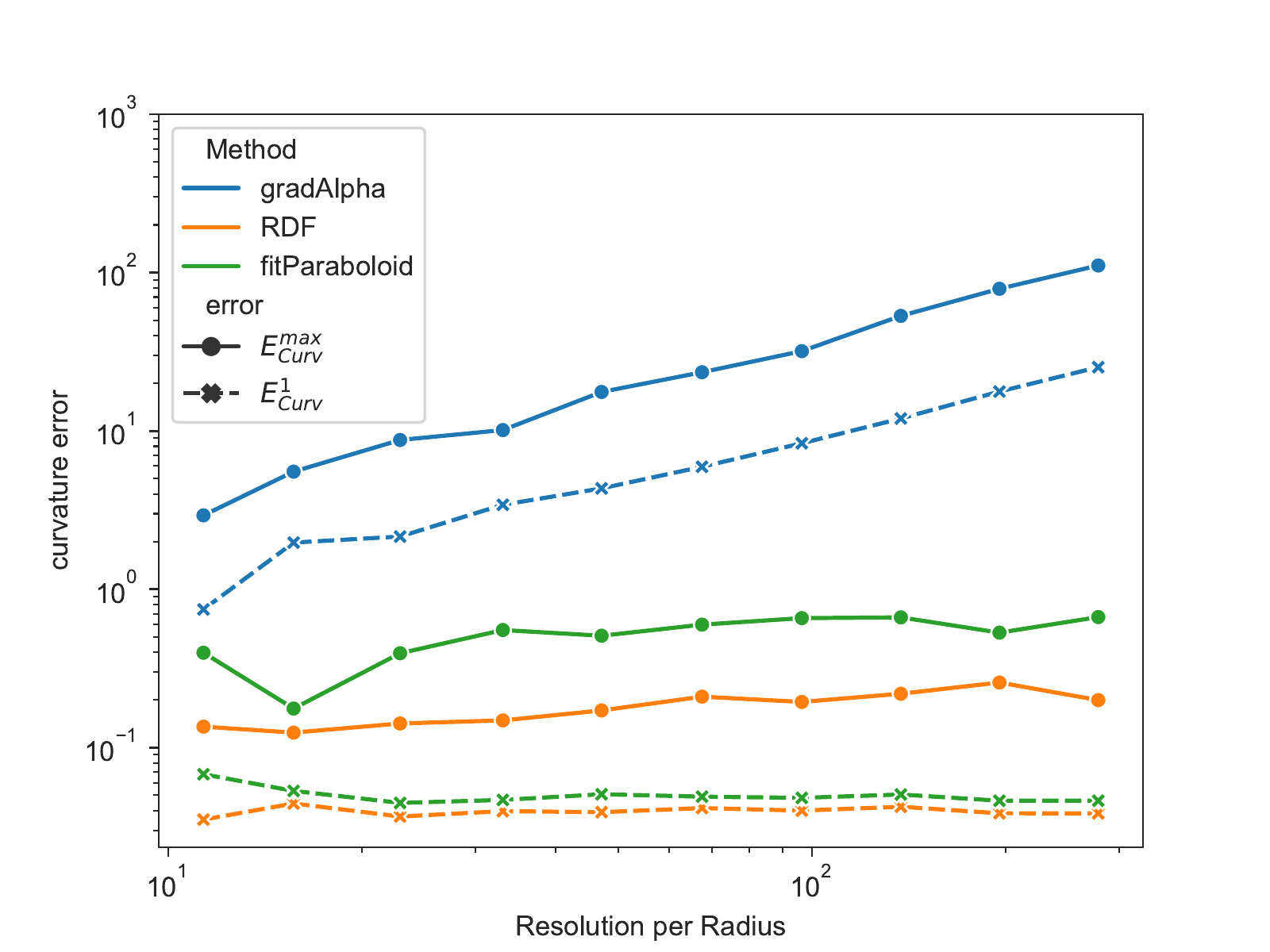}
        \caption{Tetrahedral} \label{fig:curv2d_tri}
    \end{subfigure}
    \caption{Curvature of a circle} \label{fig:curv2d}
\end{figure}

The first test case is the static reconstruction of a circle. For this, a domain with the dimensions 2m$\times$2m is created and a circular liquid region of radius 0.5 m is initialised. To test the proposed models, the cell type and resolution of the mesh are varied. The results are shown in Fig. \ref{fig:curv2d}.

The difficulty in modeling surface tension is underlined by the fact that the results vary by five orders of magnitude from the most accurate model, HFM, to the least accurate model, gradAlpha, in Fig. \ref{fig:curv2d_hex}. In this test, only the HFM model shows a converging behavior which flattens with refinement. This can be explained by the imperfect initialization of the circle as also mentioned by Coquerelle and Glockner \cite{Coquerelle.2016}. The fitParaboloid and RDF models show zero-order convergence which is an improvement in curvature computation of up to 1 to 2 orders of magnitude compared to the current standard model gradAlpha. Unfortunately, it is not possible to apply the HFM on unstructured grids which is why consequently only three models can be compared on the tetrahedral prism mesh shown in Fig. \ref{fig:curv2d_tri}. As on structured grids the new models are able to achieve up to 1 to 2 orders of magnitude more accurate results.

\begin{figure}
    \begin{subfigure}{0.5\textwidth}
        \includegraphics[width=\linewidth]{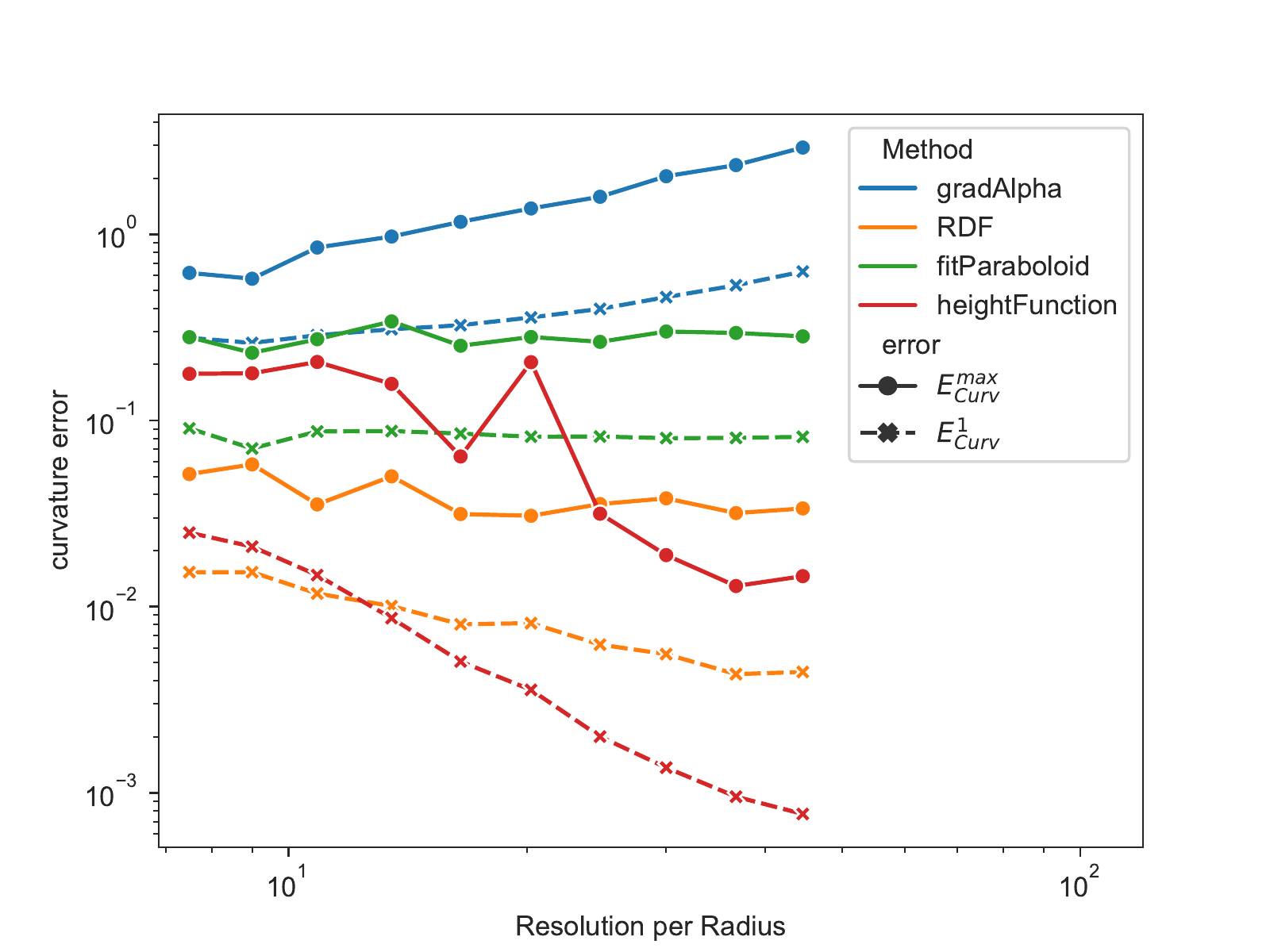}
        \caption{Hexahedral grid} \label{fig:curv3d_hex}
        \end{subfigure}
        \hspace*{\fill} 
    \begin{subfigure}{0.5\textwidth}
        \includegraphics[width=\linewidth]{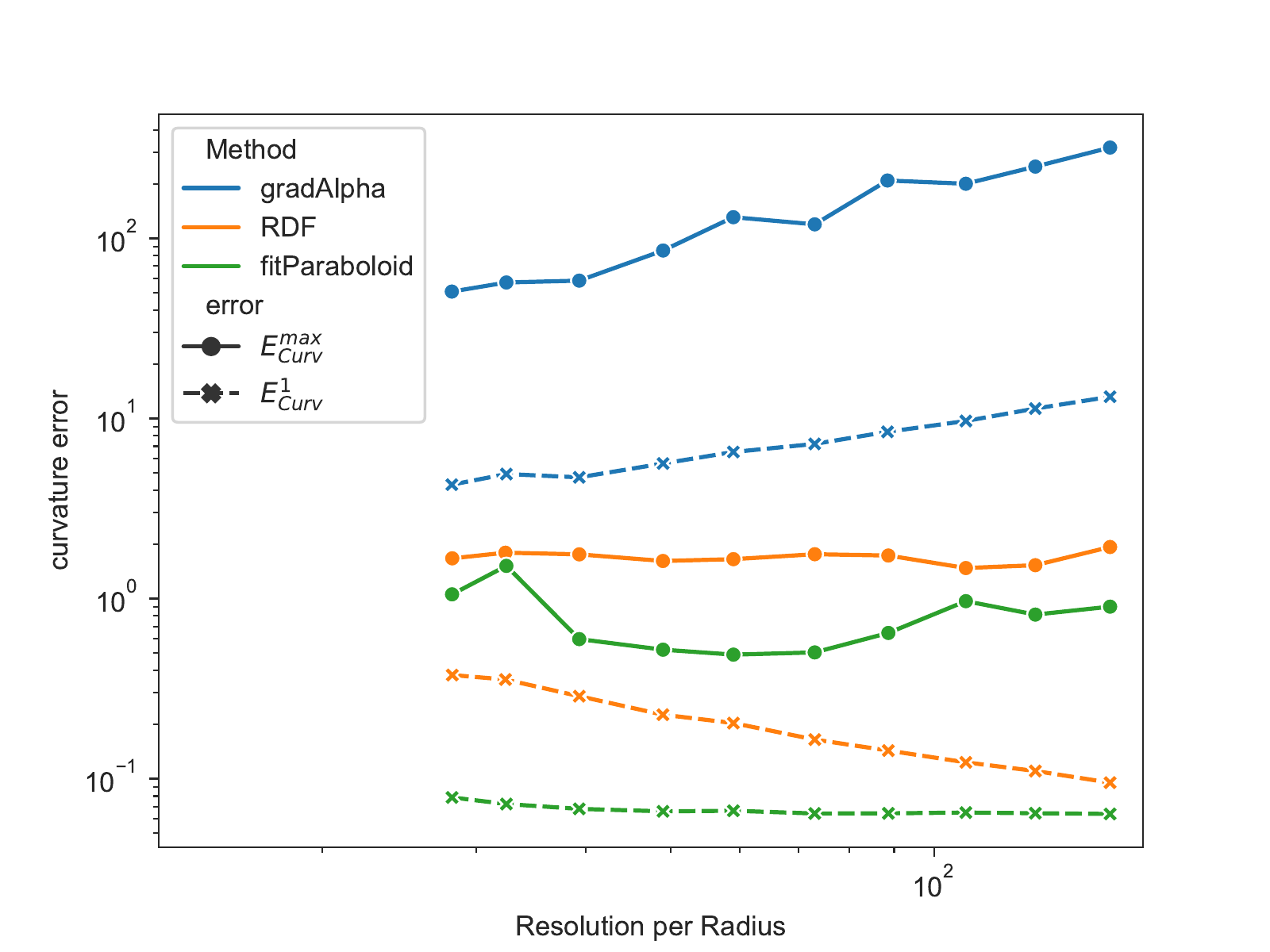}
        \caption{Tetrahedral grid} \label{fig:curv3d_tri}
    \end{subfigure}
    \caption{Curvature of a sphere} \label{fig:curv3d}
\end{figure}

The reconstruction of a sphere of radius 1m on a 2m$\times$2m domain gives a similar picture as shown in Fig.~\ref{fig:curv3d}. Again, the HFM is the most accurate method followed by RDF, fitParaboloid and finally gradAlpha.

\subsubsection{Curvature of a disc for various contact angles}

In this test case the accuracy of the curvature computation with the presence of a boundary is simulated. To test the implementation, a 4m$\times$2m domain is created and multiple circles cutting the domain boundary with angles 15, 30 , 45 ,60 and 75 degrees are initialized. As in the reconstruction test case the resolution and grid type are varied in addition to the contact angle at the boundary.

\begin{figure}
    \begin{subfigure}{0.5\textwidth}
        \includegraphics[width=\linewidth]{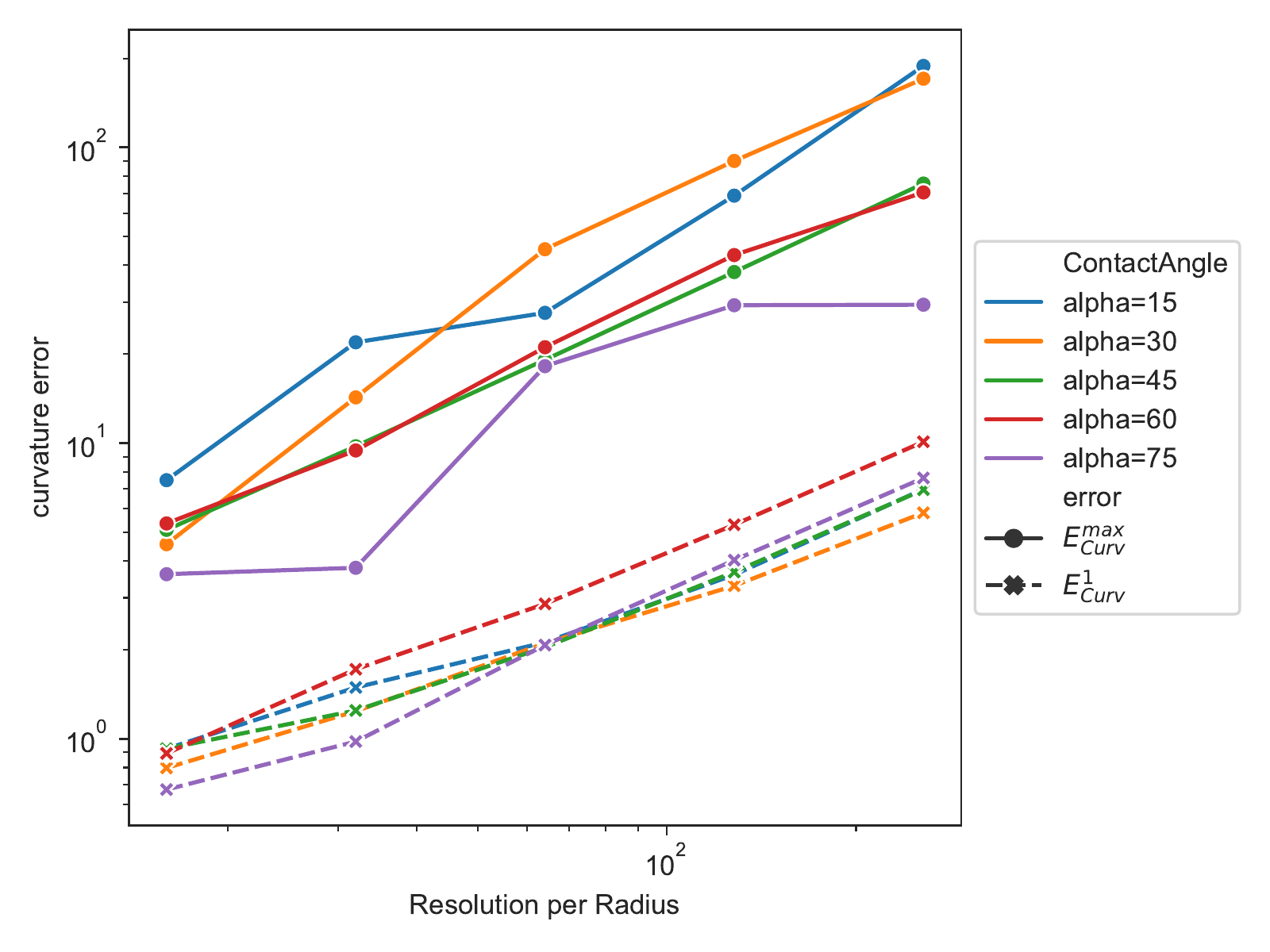}
        \caption{gradAlpha - hexahedral grid} \label{fig:curv2d_wall_gradAlpha_hex}
        \end{subfigure}
        \hspace*{\fill} 
    \begin{subfigure}{0.5\textwidth}
        \includegraphics[width=\linewidth]{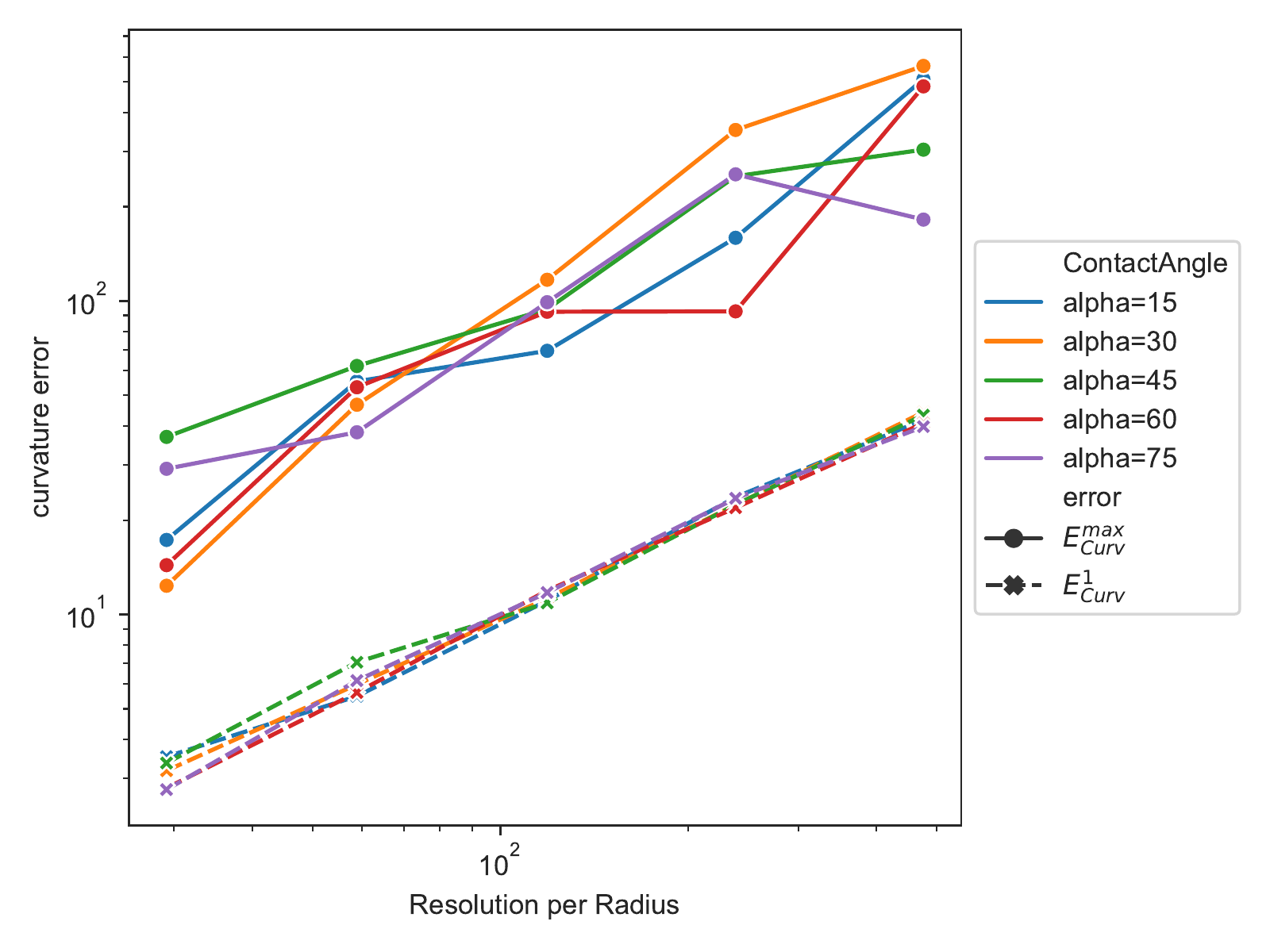}
        \caption{gradAlpha - tetrahedral grid} \label{fig:curv2d_wall_gradAlpha_tri}
    \end{subfigure}
    \begin{subfigure}{0.5\textwidth}
        \includegraphics[width=\linewidth]{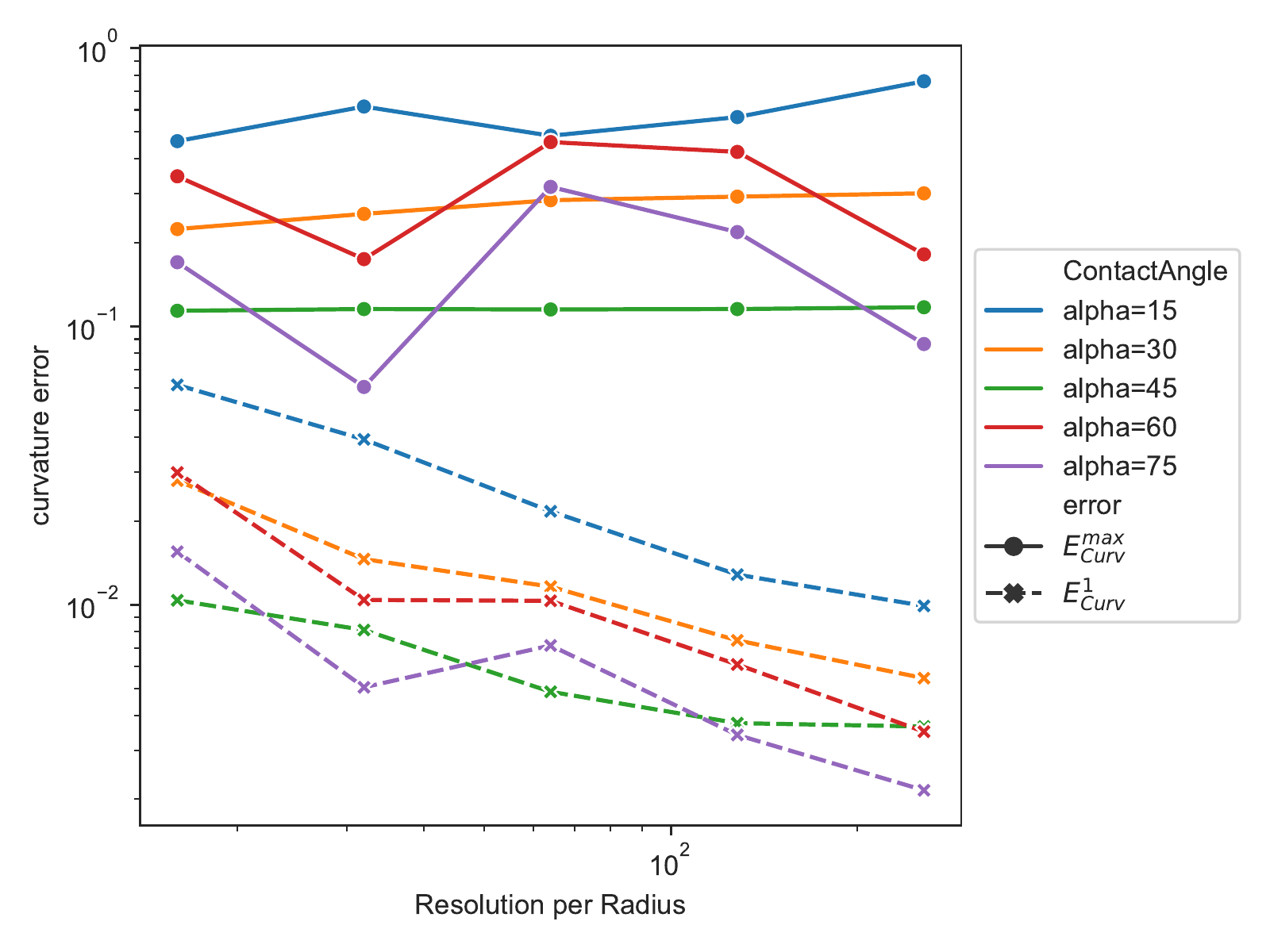}
        \caption{RDF - hexahedral grid} \label{fig:curv2d_wall_RDF_hex}
        \end{subfigure}
        \hspace*{\fill} 
    \begin{subfigure}{0.5\textwidth}
        \includegraphics[width=\linewidth]{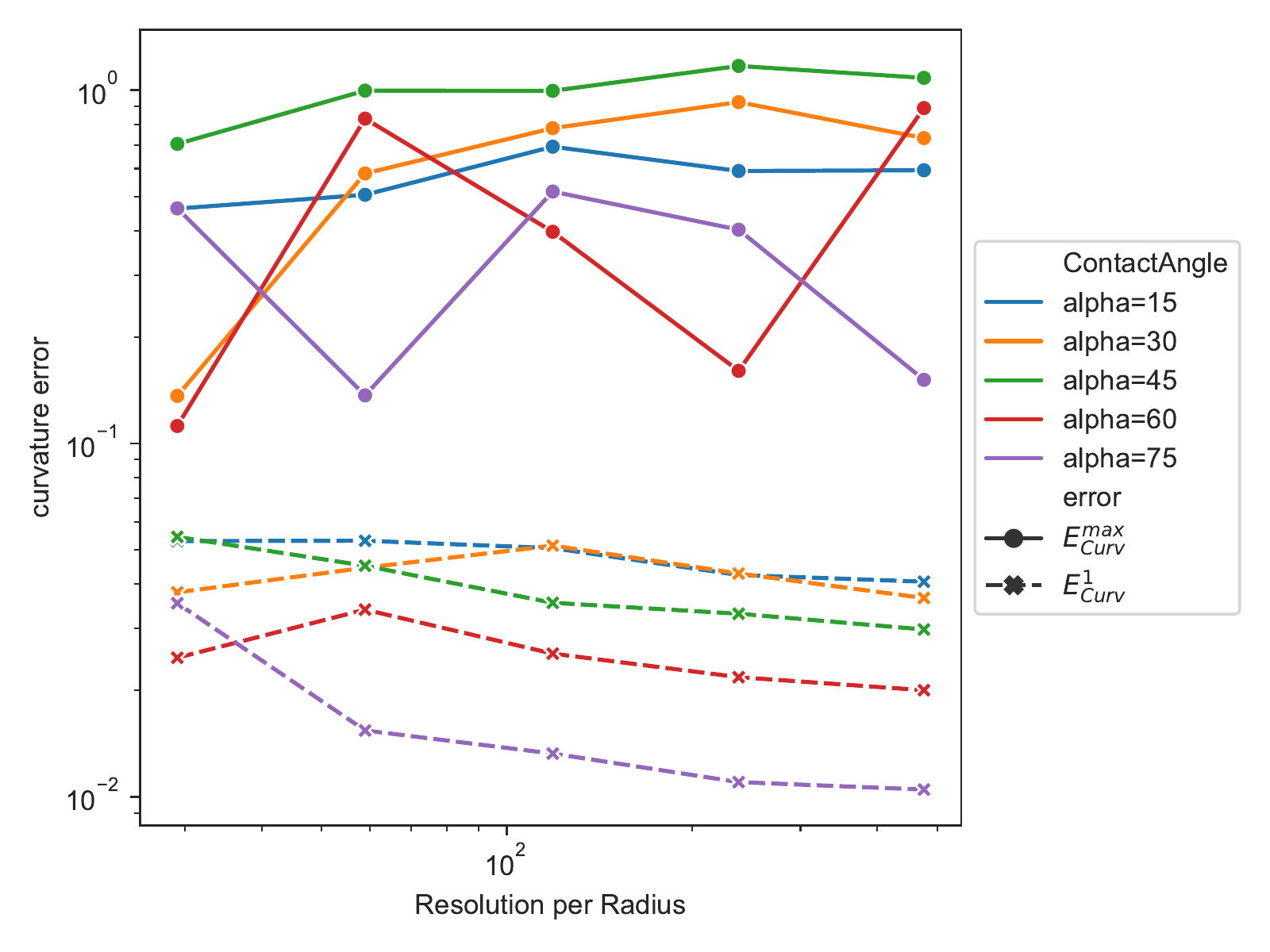}
        \caption{RDF - tetrahedral grid} \label{fig:curv2d_wall_RDF_tri}
    \end{subfigure}
    \begin{subfigure}{0.5\textwidth}
        \includegraphics[width=\linewidth]{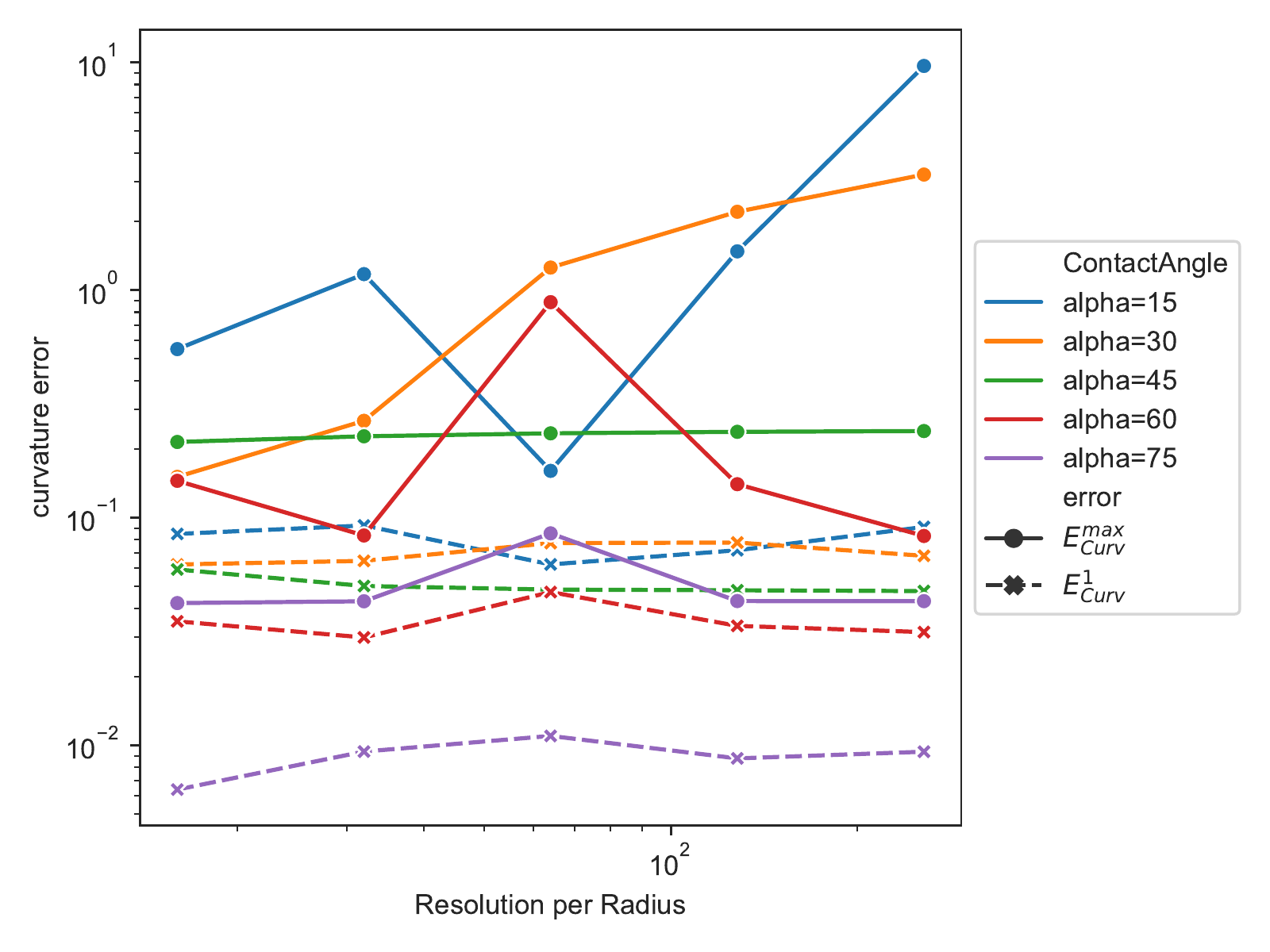}
        \caption{fitParaboloid - hexahedral grid} \label{fig:curv2d_wall_fitParaboloid_hex}
        \end{subfigure}
        \hspace*{\fill} 
    \begin{subfigure}{0.5\textwidth}
        \includegraphics[width=\linewidth]{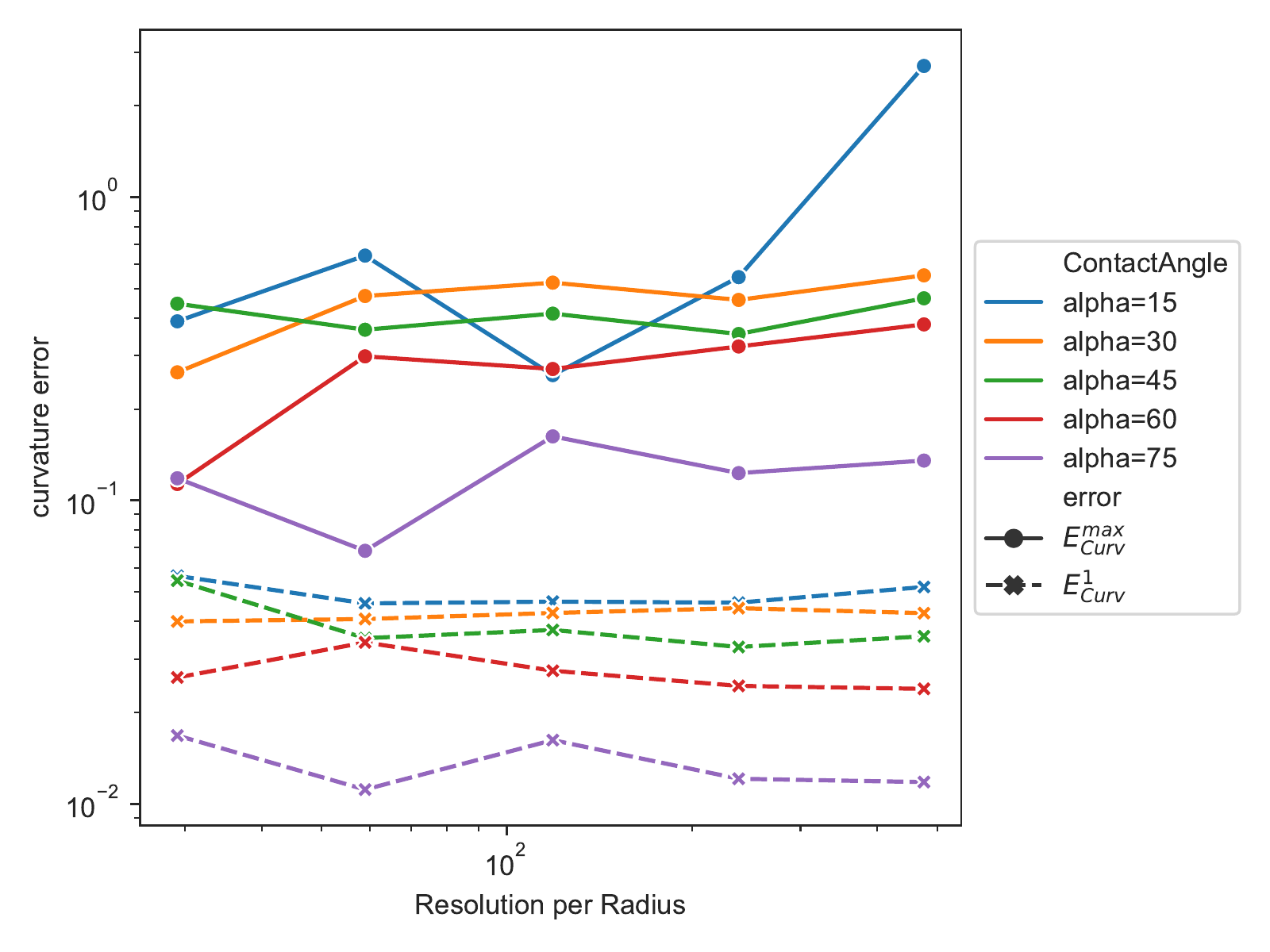}
        \caption{fitParaboloid - tetrahedral grid} \label{fig:curv2d_wall_fitParaboloid_tri}
    \end{subfigure}
    \caption{Curvature error for different contact angles} \label{fig:curv2d_wall}
\end{figure}

This test case gives an impression of the accuracy of the contact angle implementation. The curvature error (see Figs. \ref{fig:curv2d_wall_gradAlpha_hex} and \ref{fig:curv2d_wall_gradAlpha_tri}) shows a diverging behaviour as in the previous test case with the maximum error being slightly larger for smaller contact angles. This increase in error for lower contact angles is most pronounced for the RDF and fitParaboloid model as depicted in Fig. \ref{fig:curv2d_wall_RDF_hex} to \ref{fig:curv2d_wall_fitParaboloid_tri}. 

\subsubsection{Translating circle}

\begin{figure}
    \begin{subfigure}{0.5\textwidth}
        \includegraphics[width=\linewidth]{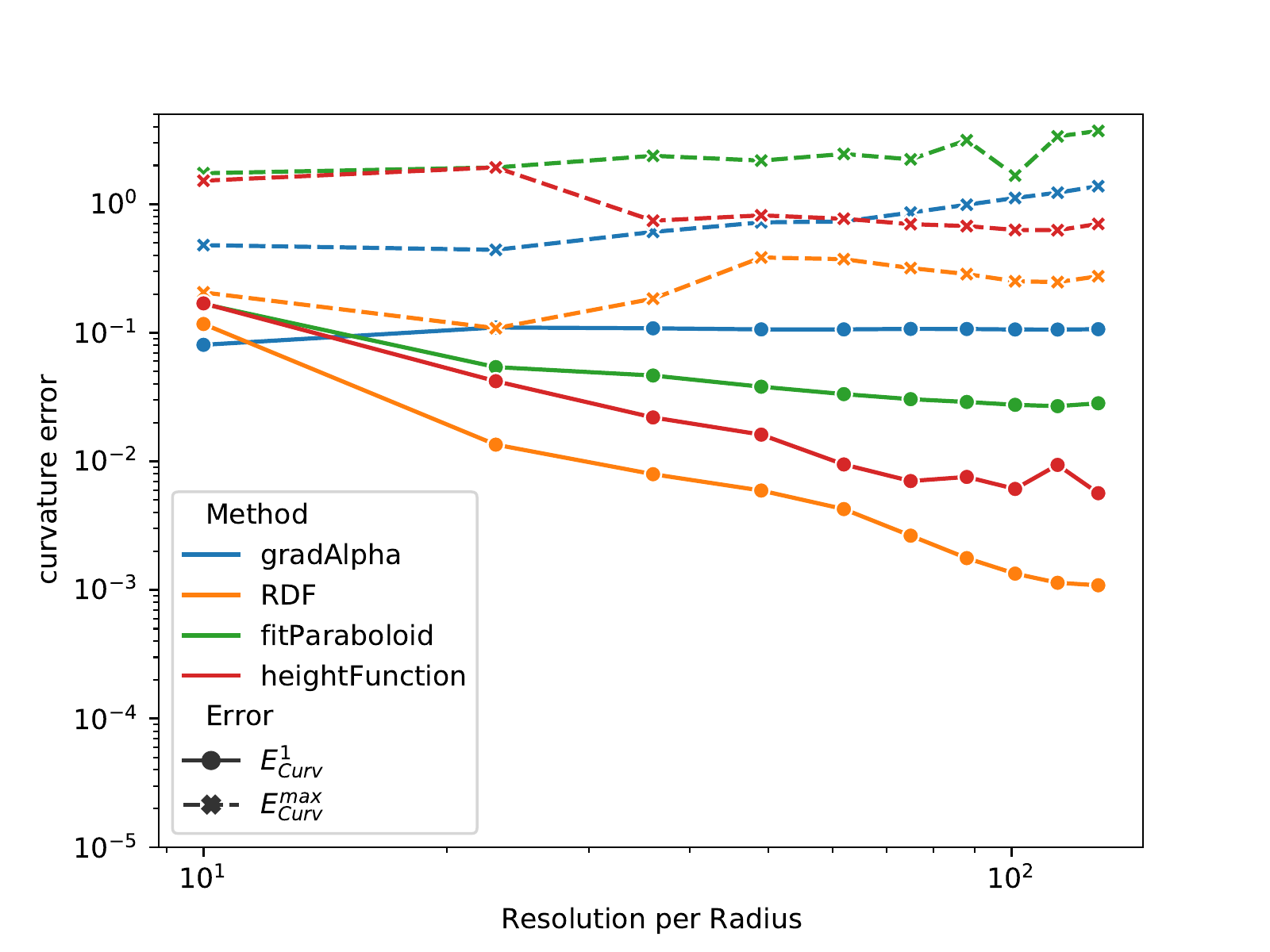}
        \caption{Hexahedral grids} \label{fig:advectCircle_hex}
        \end{subfigure}
        \hspace*{\fill} 
    \begin{subfigure}{0.5\textwidth}
        \includegraphics[width=\linewidth]{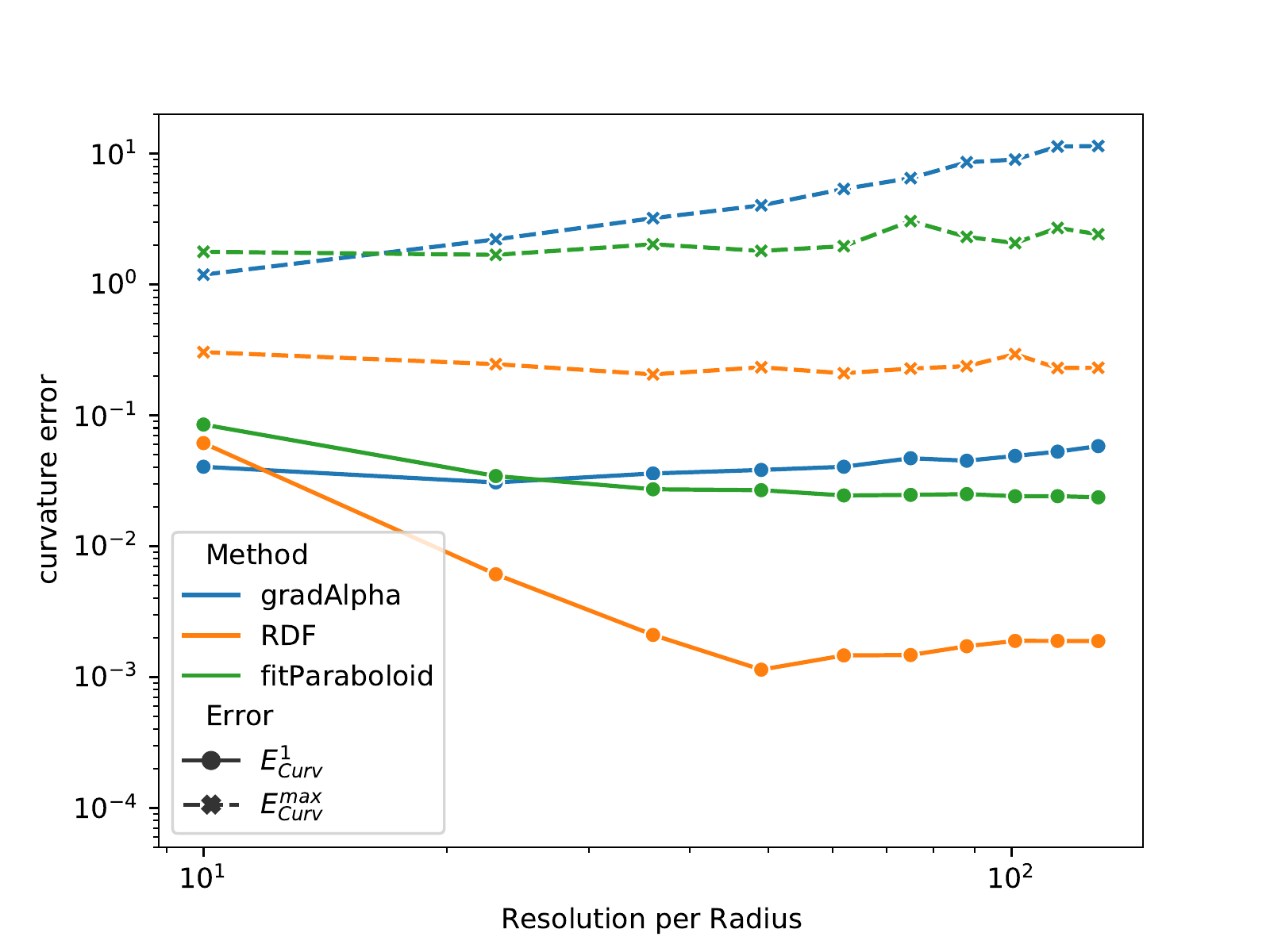}
        \caption{Tetrahedral grids} \label{fig:advectCircle_tri}
    \end{subfigure}
    \caption{Advection of a circle} \label{fig:advectCircle}
\end{figure}

So far we tested the capabilities of the different curvature models for a static interface configuration. The next step is to test them in combination with a flow solver including the errors in pressure, velocity and advection. The most basic test case is the pure advection of a circle in a spatially and temporally constant flow. The circle moves with the same velocity as the surrounding gas but the pressure inside the bubble is increased due to the Young-Laplace law. The channel has a dimension of 4 m times 1 m with a background velocity of 1 $m/s$. The gas and liquid are assumed to have identical density and kinematic viscosity, 1000 $kg/m^3$ and $3.333 \cdot 10^{-5} \textrm{ } m/s^2$, respectively. We compare the time-averaged deviation of curvature and the maximal curvature error for hexahedral and tetrahedral grids with different resolution using the proposed methods. Fig \ref{fig:advectCircle_hex} shows the two curvature error measures as functions of grid resolution. As in the static reconstruction test case the most accurate results are achieved by the HFM but the maximum error does not seem to converge which was also observed by Popinet \cite{Popinet.2009}. As in the previous test case the implemented models are significantly more accurate than the standard curvature model gradAlpha. On unstructured meshes, we see the same trend, that fitParaboloid and in particular RDF are significantly more accurate than the standard OpenFOAM model.

\subsubsection{Sine wave}

\begin{figure}
    \begin{subfigure}{0.5\textwidth}
        \includegraphics[width=\linewidth]{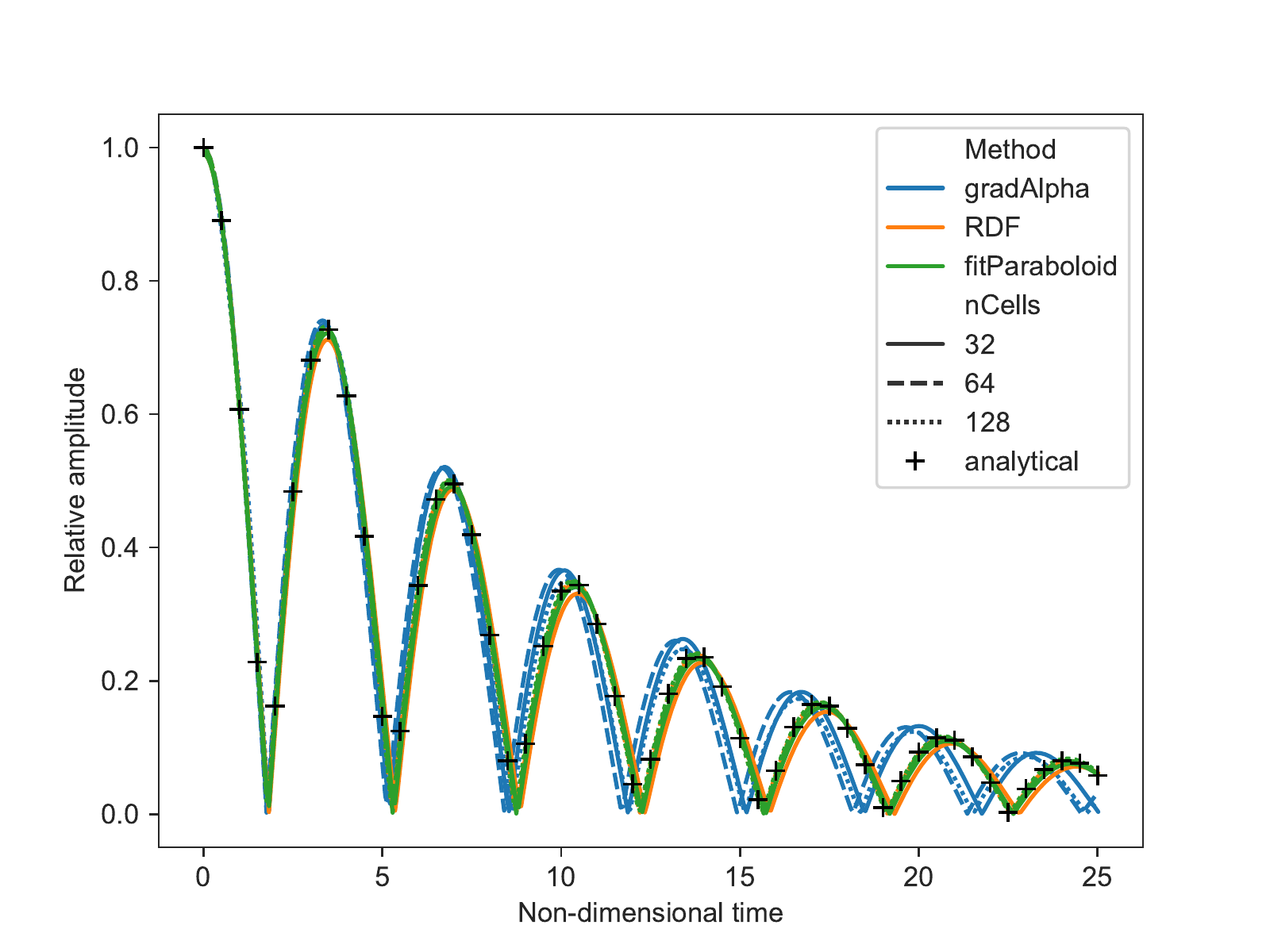}
        \caption{Geometric VoF (isoAdector)} \label{fig:sinWaveHex_plicRDF}
        \end{subfigure}
        \hspace*{\fill} 
    \begin{subfigure}{0.5\textwidth}
        \includegraphics[width=\linewidth]{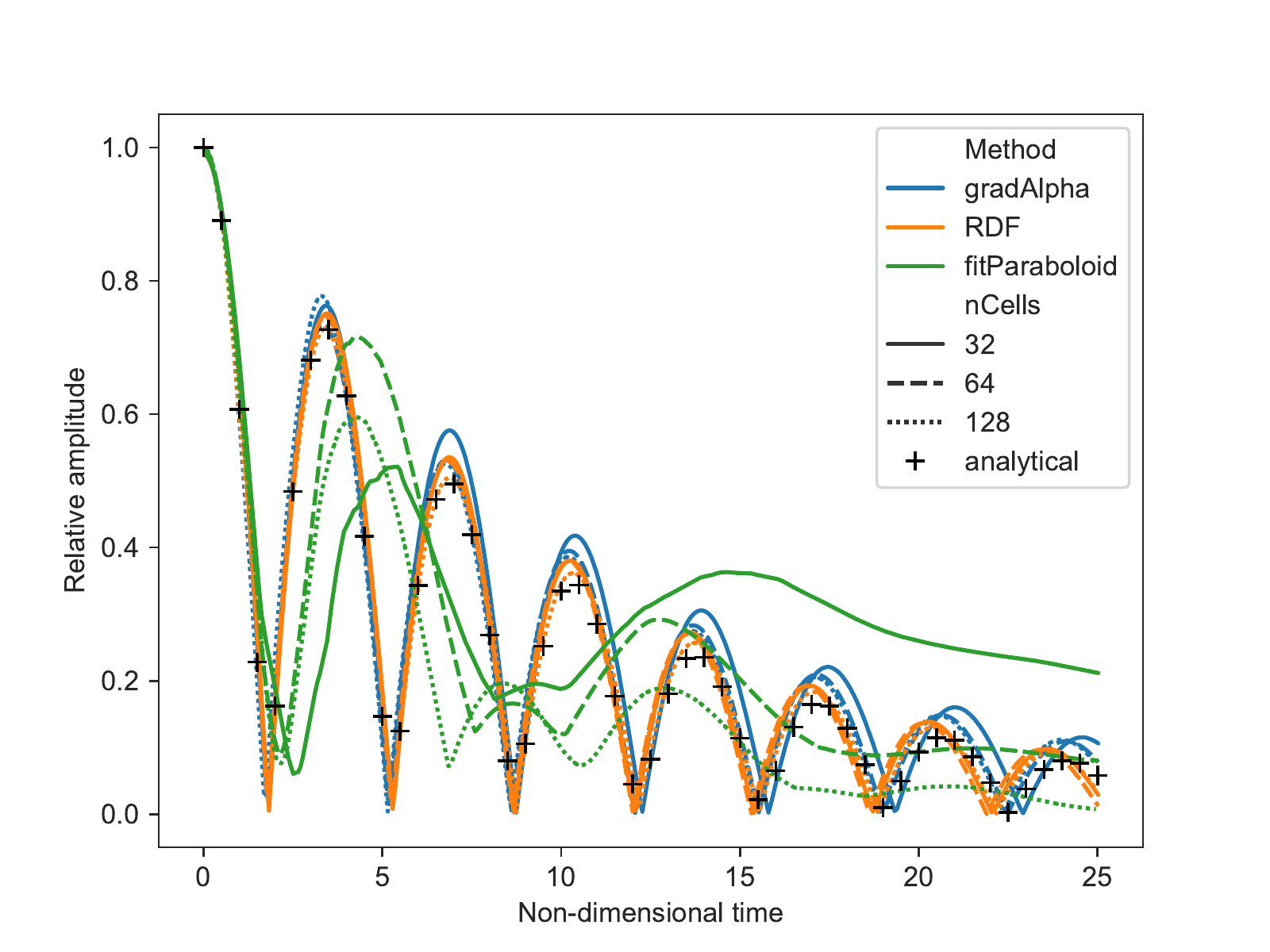}
        \caption{Colour function VoF (MULES)} \label{fig:sinWaveHex_isoSurface}
    \end{subfigure}
    \caption{Sine wave: Hexahedral grids} \label{fig:sinWaveHex}
\end{figure}

The next step is the verification of the models with the analytical solution of Prospretti \cite{Prosperetti.1981}. He found an analytical solution for the movement of a cosine wave including the effect of viscosity and surface tension. The main differences to the previous test case are that it includes the effect of the boundaries and that the curvature is not constant over the surface. Our domain and fluid properties are identical to the ones proposed by Popinet \cite{Popinet.2009}. As in the previous benchmarks, the models are compared for different grid types, grid resolutions and interface advection methods. Fig. \ref{fig:sinWaveHex_plicRDF} shows the evolution of maximum height of the sine wave for the structured grids in combination with the geometric VoF method. The RDF and fitParaboloid methods are able to capture the amplitude of the frequency with good accuracy. Only the gradAlpha method deviates substantially from the analytical solution. The HFM method is not shown since it is not able to handle contact angles in our current implementation. In Fig. \ref{fig:sinWaveHex_isoSurface} the geometric VoF method is replaced with the standard colour function VoF method. We observed that only the RDF method is capable of an accurate representation of the analytical function. The comparison of both advection methods with the available surface tension models reveals that the geometric VoF method is more accurate with the implemented surface tension methods. 

\begin{figure}
    \centering
    \includegraphics[width=0.5\linewidth]{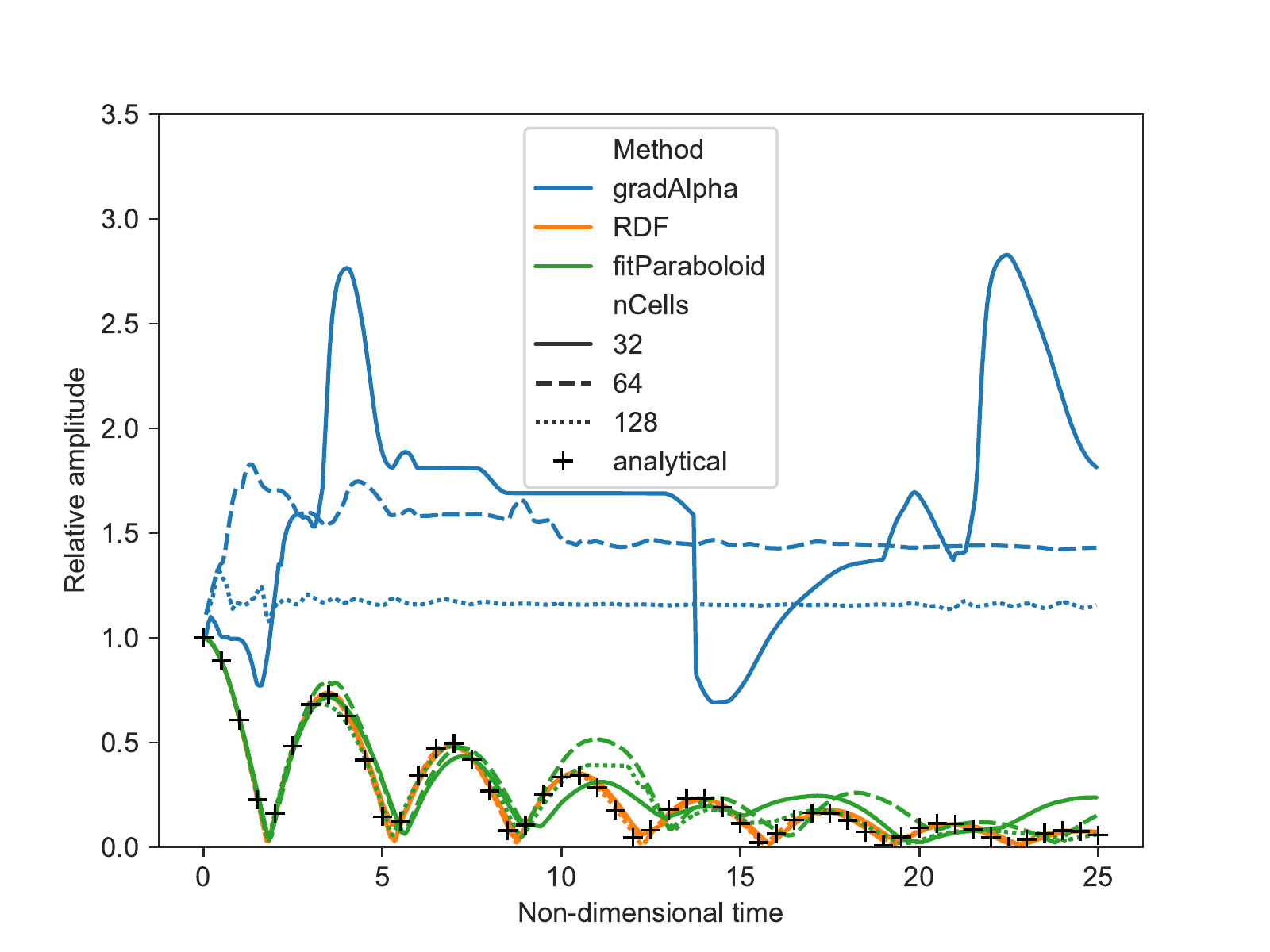}
    \caption{Sine wave: Tetrahedral grids} \label{fig:sinWaveTri_plicRDF}
\end{figure}

The combination of geometric VoF shows also good accuracy on unstructured grids which is shown in Fig. \ref{fig:sinWaveTri_plicRDF}. The gradAlpha and fitParaboloid methods show a significantly reduced accuracy compared to the structured grid. However, the fitParaboloid method at least qualitatively shows behaviour similar to the analytical solution in contrast to gradAlpha.

\newpage
\section{Conclusions and future work}

The OpenFOAM based code framework, TwoPhaseFlow, offers new phase change and surface tension models. The phase change module offers three phase change models denoted explicitGrad, implicitGrad and Schrage. The implementations have been validated using simple analytical benchmark cases from the literature. Currently, three surface tension models are available: The Height Function Method, the Reconstructed Distance Function method and the FitParaboloid method. These new models are validated with surface tension benchmarks and show a reduction of the spurious currents by more than an order of magnitude depending on the model choice. The library offers three solvers of which the most general is able to simulate compressible two phase flow including the effects of the phase change and surface tension. All models work with both the isoAdvector geometric VoF method as well as the MULES method for interface advection. The TwoPhaseFlowLib framework capitalizing on the runtime selection mechanism of OpenFOAM allowing easy implementation and verification of new models. The framework is released under the GPL v3 and the source code is publicly-available in a software repository \cite{TwoPhaseFlow}.

\section*{Acknowledgement}

This work was supported by German Aerospace Center - DLR. JR acknowledges support from Independent Research Fund Denmark (Grant-ID: 9063-00018B). The authors would like to thank Grega Belsak, Lionel Gamet, Niklas Weber, Marco Scala and Christoph Wilms  for testing the library and reporting numerous bugs.

\clearpage

\appendix

\section{Thermodynamic model}
\label{app:thermo}

In this paper only phase change models caused by temperature gradients are implemented but for numerous engineering application concentration based phase change is crucial. This framework tries to offer a possibility to simplify the implementation of the concentration based phase change. The solver \texttt{multiRegionPhaseChangeFlow} is able to account for mixture of different species in each phase by utilizing a similar thermodynamic framework to \texttt{icoReactingMultiphaseInterFoam}. The activation of the multi components in phase is done by switching a keyword.

\noindent
\begin{small}
\begin{minipage}{\linewidth}
\begin{verbatim}
phases (liquid gas);

liquid
{
    type purePhaseModel;
}

gas
{
    type multiComponentPhaseModel;
    Sc              0.7;
    residualAlpha   1e-3;
}
\end{verbatim}
\end{minipage}
\end{small}

The base of \texttt{multiComponentPhaseModel} is \texttt{rhoReactionThermo} and therefore gives the possibility to include reaction in the future as well. In the current state, the concentration fields only affect the density of the phase and does not accounted for phase change at the interface.

\clearpage
 
\bibliographystyle{elsarticle-num}
\bibliography{citations}










\end{document}